\documentclass{article}

\usepackage[utf8]{inputenc}
\usepackage[english]{babel}
\usepackage[backend=biber,style=ieee,citestyle=numeric]{biblatex}
\usepackage{authblk}
\usepackage{csvsimple}
\usepackage{float}
\usepackage{hyperref}
\usepackage{csquotes}
\usepackage{booktabs,array}
\usepackage{nameref}
\usepackage{graphicx}
\usepackage{float}
\usepackage{caption}
\usepackage[dvipsnames,svgnames,usenames,table]{xcolor}

\usepackage[toc,acronym,translate=babel,acronymlists={hidden},automake]{glossaries-extra}
\GlsXtrEnablePreLocationTag{~-~page:~}{~-~pages:~}
\glssetcategoryattribute{general}{textformat}{emph}
\glssetcategoryattribute{general}{glossnamefont}{emph}
\glssetcategoryattribute{acronym}{glossnamefont}{textbf}
\newglossarystyle{glsstylelong}{
    \setglossarystyle{long}
    
}
\setabbreviationstyle[acronym]{short-long}
\setabbreviationstyle[main]{short-long}

\newglossary[glhidden]{hidden}{glhidden}{glhiddenin}{Hidden Glossary}

\makeglossaries
\loadglsentries{anc/gomes2023influencers.gls}

\usepackage[noabbrev,capitalise]{cleveref}

\newrobustcmd*{\citefullauthor}{\AtNextCite{\DeclareNameAlias{labelname}{given-family}}\citeauthor}
\title{Recommending Influencers to Merchants\\using Matching Game Algorithm}
\author[1]{Gomes, J.M.}
\author[1]{Dias, L.A.V.}
\affil[1]{Instituto Tecnológico de Aeronáutica - ITA}

\begin{document}

\maketitle

\section*{Abstract}


The goal of this work was to apply the ``\textbf{Gale-Shapley}'' algorithm to a real-world problem. We analyzed the pairing of influencers with merchants, and after a detailed specification of the variables involved, we conducted experiments to observe the validity of the approach.


We conducted an analysis of the problem of aligning the interests of merchants to have digital influencers promote their products and services. We propose applying the matching algorithm approach to address this issue.


We demonstrate that it is possible to apply the algorithm and still achieve corporate objectives by translating performance indicators into the desired ranking of influencers and product campaigns to be advertised by merchants.

\section*{Introduction}



The ``Digital Influencer'' is a phenomenon that emerged around 2015\footnote{During the same period, the term ``Content Creators'' was also used} and represents an evolution of ``bloggers'' and ``vloggers''. Based on the reach of these individuals, they have taken on roles as opinion leaders and have been enlisted by companies for merchandising projects \cite{karhawi2017influenciadores}.



Studies not directly related to recommendation systems address the clustering of consumers according to their preferences, defining profiles of consumption, attention, personal, or social objectives \cite{kanavosLargeScaleProduct2018}. ``\textbf{Gale-Shapley}'' focused on the formation of stable pairs between any two groups, meaning the formed pairs are within the preference lists of each individual \cite{gale2013college}, and no pair is formed with individuals outside each person's personal list.


Several variations of the algorithm have been created, such as:


\begin{enumerate}
    \item Heterogeneous players: for example, men and women in marriage;
    \item Homogeneous players: roommates;
    \item $1-1$ relationships: again, marriage or roommates;
    \item $m-1$ relationships: residents and hospitals or students and projects \cite{roth1985college}.
\end{enumerate}


In this work, we approach a recommendation system as a matching game where a finite number of influencers (those who recommend) and an also finite number of merchants and industrialists with their services and products need to be paired. Each influencer is willing to recommend items to their clients (in commerce and industry), who in turn have a limited supply of advertisements. Influencers, merchants, and industrialists are heterogeneous, but we can assume the recommendation itself as a homogeneous element. Each agent (influencer or consumer) derives utility from the attention received or given with additive objective functions. It is an $n-to-m$ two-sided model, with attention vectors, one for each recommendation, and an allocation of recommendations to listeners or readers, so that the demand of each is satisfied, and the number of recommendations made by the influencer does not exceed the available attention.

We applied the algorithm developed based on the work of \citefullauthor{gale2013college} \cite{gale2013college} with the revisions by \citefullauthor{roth1985college} \cite{roth1985college,roth1992two}, using the implementation by \citefullauthor{wilde2020matching} \cite{wilde2020matching} as a reference. This was done in a proof of concept where we aimed to achieve the optimal matching between influencers, merchants, and their products and services.

\subsection*{Problem statement}\label{subsec:definicao}


Let $F = \{ f_1, f_2, \cdots, f_n \}$ be a set of influencers, $P = \{ p_1, p_2, \cdots, p_n \}$ a list of available products, and $V = \{ v_1, v_2, \cdots, v_3 \}$ a list of merchants selling these products.


Each influencer $f_i$ provides a preference list with items from $P$ (see \cref{tab:influencers}). If product $p_j$ appears in the list of $f_i$, then $f_i$ considers $p_j$ desirable. We denote $D_i$ as the set of products desired by $f_i$.

\begin{table}[ht]\scriptsize
    \centering
    \begin{tabular}{c r l l l l}
        \toprule \bfseries Id $f_i$ & \bfseries Rep. $r(f_i)$ & \bfseries $p_1$ & \bfseries $p_2$ & $\cdots$ & \bfseries $p_j$ \\
        \midrule
        \begin{filecontents*}{anc/stamps-gt-1-dataInfluenciadoresRankProdutos.csv}
nome,rank,0,1,2,3,4,5,6,7,8,9,10,11,12,13,14,15,16,17,18
296387a17f2b,148.0,437441,11534,498323,332334,127099,15168,573,822143,822139,682879,12436,13121,16720,680731,834004,10028,16051,10830,377381
64c6cf346842,62.0,9976,572953,0,0,0,0,0,0,0,0,0,0,0,0,0,0,0,0,0
89f3d85b7645,57.0,367959,621987,820505,10257,11888,0,0,0,0,0,0,0,0,0,0,0,0,0,0
89cca91ef90c,49.0,850416,0,0,0,0,0,0,0,0,0,0,0,0,0,0,0,0,0,0
1a935a652ab0,18.0,673576,0,0,0,0,0,0,0,0,0,0,0,0,0,0,0,0,0,0
86e29670dfa9,3.0,228771,0,0,0,0,0,0,0,0,0,0,0,0,0,0,0,0,0,0
f239cf761b93,3.0,17513,0,0,0,0,0,0,0,0,0,0,0,0,0,0,0,0,0,0
6e8f8d1866ce,2.0,517530,0,0,0,0,0,0,0,0,0,0,0,0,0,0,0,0,0,0
8fc36313725a,2.0,501342,0,0,0,0,0,0,0,0,0,0,0,0,0,0,0,0,0,0
\end{filecontents*}
\csvreader[late after line=\\\hline, late after last line=\\, head to column names, filter={\value{csvrow}<5}]
            {anc/stamps-gt-1-dataInfluenciadoresRankProdutos.csv}
            {}
            {\nome & \rank & \0 & \1 & $\cdots$ & -}
        \bottomrule
    \end{tabular}
    \caption{Sample of the list of influencers and their preferences}
    \label{tab:influencers}
\normalsize\end{table}

\begin{table}[ht]\scriptsize
    \centering
    \begin{tabular}{c r l l l l}
        \toprule \bfseries Id $f_i$ & \bfseries Rep. $r(f_i)$ & \bfseries $p_1$ & \bfseries $p_2$ & $\cdots$ & \bfseries $p_j$ \\
        \midrule
        \begin{filecontents*}{anc/stamps-fm-1-dataInfluenciadoresRankProdutos.csv}
nome,rank,0,1,2,3,4,5,6,7,8,9,10,11,12,13,14,15,16,17,18
296387a17f2b,79.0,437441,11534,498323,332334,127099,15168,573,822143,822139,682879,12436,13121,16720,680731,834004,10028,16051,10830,377381
64c6cf346842,44.0,9976,572953,0,0,0,0,0,0,0,0,0,0,0,0,0,0,0,0,0
89f3d85b7645,26.0,367959,621987,820505,10257,11888,0,0,0,0,0,0,0,0,0,0,0,0,0,0
89cca91ef90c,14.0,850416,0,0,0,0,0,0,0,0,0,0,0,0,0,0,0,0,0,0
1a935a652ab0,8.0,673576,0,0,0,0,0,0,0,0,0,0,0,0,0,0,0,0,0,0
f239cf761b93,3.0,17513,0,0,0,0,0,0,0,0,0,0,0,0,0,0,0,0,0,0
8fc36313725a,3.0,501342,0,0,0,0,0,0,0,0,0,0,0,0,0,0,0,0,0,0
86e29670dfa9,2.0,228771,0,0,0,0,0,0,0,0,0,0,0,0,0,0,0,0,0,0
6e8f8d1866ce,2.0,517530,0,0,0,0,0,0,0,0,0,0,0,0,0,0,0,0,0,0
\end{filecontents*}
\csvreader[late after line=\\\hline, late after last line=\\, head to column names, filter={\value{csvrow}<5}]
            {anc/stamps-fm-1-dataInfluenciadoresRankProdutos.csv}
            {}
            {\nome & \rank & \0 & \1 & $\cdots$ & -}
        \bottomrule
    \end{tabular}
    \caption{Sample of the list of influencers and their preferences classified by Average Purchases}
    \label{tab:influencersmc}
\normalsize\end{table}

Each merchant $v_k$ advertises a list of products $P_k$ where $P_1, P_2, \cdots, P_l$ segments $P$ (see \cref{tab:merchants}). For each product announced in $P_k$, the merchant indicates how many influencers they want to allocate. Let $M_k = \{ f_i \in F: P_k \cap D_i \ne \emptyset \}$ be the set $M_k$ of influencers who wish to advertise the products offered by merchant $v_k$. Consider $r(f_i)$ as the reputation of the influencer. Thus, for all $p_j \in P_k: \nu $, we represent $\nu_k^j$ as the preferences of $v_k$ for product $p_j$ - obtained by excluding all influencers $f_i$ not interested in $p_j$ in the order dictated by $r(f_i)$. At the same time, while each merchant $v_k$ is limited to a quantity $q_k$ of influencers they can accept (see \cref{tab:merchants}), each product $p_j$ also has a limit $l_j$ on how many influencers it can be offered to.

\begin{table}[ht]\scriptsize
    \centering
    \begin{tabular}{c r}
        \toprule \bfseries Id $v_k$ & \bfseries Quota $q_k$ \\
        \midrule
        \begin{filecontents*}{anc/stamps-fm-2-dataAnunciantesQuota.csv}
nome,quota
43d8ac5e2f82,45
c8923e87aa1f,3
b7b3db8ade87,2
7f370a5d8046,1
6c6d30a69b40,1
9803e6f1a3ff,1
\end{filecontents*}
\csvreader[late after line=\\\hline, late after last line=\\, head to column names, filter={\value{csvrow}<5}]
            {anc/stamps-fm-2-dataAnunciantesQuota.csv}
            {}
            {\nome & \quota}
        \bottomrule
    \end{tabular}
    \caption{Sample of the list of merchants}
    \label{tab:merchants}
\normalsize\end{table}

\begin{table}[ht]\scriptsize
    \centering
    \begin{tabular}{c r c}
        \toprule \bfseries Cod $p_j$ & \bfseries Quota $l_j$ & \bfseries Com. $v_k$ \\
        \midrule
        \begin{filecontents*}{anc/stamps-fm-3-dataProdutoComercianteQuota.csv}
codigo,quota,comerciante
823128,43,43d8ac5e2f82
12586,16,43d8ac5e2f82
437441,16,43d8ac5e2f82
7976,15,43d8ac5e2f82
17150,12,43d8ac5e2f82
127099,12,43d8ac5e2f82
528334,10,43d8ac5e2f82
697758,10,43d8ac5e2f82
682879,7,43d8ac5e2f82
367959,6,43d8ac5e2f82
669202,6,43d8ac5e2f82
401312,6,43d8ac5e2f82
9976,6,43d8ac5e2f82
4629,6,43d8ac5e2f82
686796,5,43d8ac5e2f82
673576,5,43d8ac5e2f82
364781,5,43d8ac5e2f82
573,5,43d8ac5e2f82
11534,5,43d8ac5e2f82
822366,5,43d8ac5e2f82
686807,5,43d8ac5e2f82
501400,5,43d8ac5e2f82
498323,4,43d8ac5e2f82
454373,4,43d8ac5e2f82
15168,4,43d8ac5e2f82
590634,4,43d8ac5e2f82
35238,4,43d8ac5e2f82
15285,4,43d8ac5e2f82
17513,4,43d8ac5e2f82
332334,4,43d8ac5e2f82
621987,4,43d8ac5e2f82
501342,4,43d8ac5e2f82
406108,4,43d8ac5e2f82
681132,4,43d8ac5e2f82
5607,4,43d8ac5e2f82
292992,4,43d8ac5e2f82
12436,4,43d8ac5e2f82
9562,4,43d8ac5e2f82
1054,3,43d8ac5e2f82
820505,3,43d8ac5e2f82
846119,3,43d8ac5e2f82
404171,3,43d8ac5e2f82
846771,3,43d8ac5e2f82
680731,3,43d8ac5e2f82
452395,3,43d8ac5e2f82
381196,3,43d8ac5e2f82
10329,3,43d8ac5e2f82
181121,3,43d8ac5e2f82
260,3,43d8ac5e2f82
4917,3,43d8ac5e2f82
443619,3,43d8ac5e2f82
17130,3,43d8ac5e2f82
822139,3,43d8ac5e2f82
828865,3,43d8ac5e2f82
564267,3,43d8ac5e2f82
5236,3,43d8ac5e2f82
930,3,43d8ac5e2f82
822143,3,43d8ac5e2f82
17324,3,43d8ac5e2f82
838479,2,43d8ac5e2f82
834004,2,43d8ac5e2f82
328366,2,43d8ac5e2f82
1897,2,43d8ac5e2f82
621984,2,43d8ac5e2f82
406106,2,43d8ac5e2f82
5278,2,43d8ac5e2f82
34938,2,43d8ac5e2f82
173468,2,43d8ac5e2f82
847780,2,43d8ac5e2f82
74140,2,43d8ac5e2f82
443002,2,43d8ac5e2f82
652697,2,43d8ac5e2f82
10877,2,43d8ac5e2f82
818798,2,43d8ac5e2f82
4626,2,43d8ac5e2f82
356706,2,43d8ac5e2f82
452399,2,43d8ac5e2f82
666527,2,43d8ac5e2f82
328822,2,43d8ac5e2f82
498339,2,43d8ac5e2f82
152308,2,43d8ac5e2f82
16720,2,43d8ac5e2f82
213789,2,43d8ac5e2f82
404962,2,43d8ac5e2f82
13121,2,43d8ac5e2f82
38707,2,43d8ac5e2f82
826660,2,43d8ac5e2f82
848738,2,43d8ac5e2f82
406111,2,43d8ac5e2f82
597841,2,43d8ac5e2f82
610721,2,43d8ac5e2f82
444990,2,43d8ac5e2f82
818800,2,43d8ac5e2f82
12920,2,43d8ac5e2f82
582981,2,43d8ac5e2f82
44862,2,43d8ac5e2f82
427652,2,43d8ac5e2f82
14729,2,43d8ac5e2f82
229334,2,43d8ac5e2f82
648435,2,43d8ac5e2f82
10257,2,43d8ac5e2f82
12876,2,43d8ac5e2f82
498346,2,43d8ac5e2f82
649448,2,43d8ac5e2f82
8139,2,43d8ac5e2f82
846146,2,43d8ac5e2f82
17440,2,43d8ac5e2f82
522683,2,43d8ac5e2f82
7641,2,43d8ac5e2f82
65230,1,43d8ac5e2f82
822097,1,43d8ac5e2f82
809051,1,43d8ac5e2f82
230619,1,43d8ac5e2f82
12926,1,43d8ac5e2f82
9682,1,43d8ac5e2f82
389413,1,43d8ac5e2f82
830601,1,43d8ac5e2f82
10012,1,43d8ac5e2f82
468,1,43d8ac5e2f82
1027,1,43d8ac5e2f82
16248,1,43d8ac5e2f82
179882,1,43d8ac5e2f82
664272,1,43d8ac5e2f82
17514,1,43d8ac5e2f82
44311,1,43d8ac5e2f82
127102,1,43d8ac5e2f82
825738,1,43d8ac5e2f82
139085,1,43d8ac5e2f82
470838,1,43d8ac5e2f82
176330,1,43d8ac5e2f82
2268,1,43d8ac5e2f82
5087,1,43d8ac5e2f82
10830,1,43d8ac5e2f82
818775,1,43d8ac5e2f82
283017,1,43d8ac5e2f82
293060,1,43d8ac5e2f82
850416,1,43d8ac5e2f82
12695,1,43d8ac5e2f82
128542,1,43d8ac5e2f82
673900,1,43d8ac5e2f82
836821,1,43d8ac5e2f82
3795,1,43d8ac5e2f82
16272,1,43d8ac5e2f82
993,1,43d8ac5e2f82
53004,1,43d8ac5e2f82
3707,1,43d8ac5e2f82
44513,1,43d8ac5e2f82
3851,1,43d8ac5e2f82
171643,1,43d8ac5e2f82
10213,1,43d8ac5e2f82
491362,1,43d8ac5e2f82
392130,1,43d8ac5e2f82
517530,1,43d8ac5e2f82
105614,1,43d8ac5e2f82
554790,1,43d8ac5e2f82
383357,1,43d8ac5e2f82
351721,1,43d8ac5e2f82
512464,1,43d8ac5e2f82
814912,1,43d8ac5e2f82
112,1,43d8ac5e2f82
456156,1,43d8ac5e2f82
312983,1,43d8ac5e2f82
10786,1,43d8ac5e2f82
11888,1,43d8ac5e2f82
41980,1,43d8ac5e2f82
684650,1,43d8ac5e2f82
334884,1,43d8ac5e2f82
818556,1,43d8ac5e2f82
822036,1,43d8ac5e2f82
5606,1,43d8ac5e2f82
411681,1,43d8ac5e2f82
664174,1,43d8ac5e2f82
96875,1,43d8ac5e2f82
5091,1,43d8ac5e2f82
328820,1,43d8ac5e2f82
127030,1,43d8ac5e2f82
649588,1,43d8ac5e2f82
19360,1,43d8ac5e2f82
505914,1,43d8ac5e2f82
16735,1,43d8ac5e2f82
219229,1,43d8ac5e2f82
680710,1,43d8ac5e2f82
682875,1,43d8ac5e2f82
15178,1,43d8ac5e2f82
151936,1,43d8ac5e2f82
338366,1,43d8ac5e2f82
14193,1,43d8ac5e2f82
243,1,43d8ac5e2f82
518439,1,43d8ac5e2f82
87999,1,43d8ac5e2f82
42539,1,43d8ac5e2f82
610098,1,43d8ac5e2f82
322192,1,43d8ac5e2f82
17058,1,43d8ac5e2f82
16774,1,43d8ac5e2f82
261570,1,43d8ac5e2f82
4891,1,43d8ac5e2f82
820244,1,43d8ac5e2f82
176,1,43d8ac5e2f82
972,1,43d8ac5e2f82
492816,1,43d8ac5e2f82
1139,1,43d8ac5e2f82
362101,1,43d8ac5e2f82
662676,1,43d8ac5e2f82
349708,1,43d8ac5e2f82
35673,1,43d8ac5e2f82
16268,1,43d8ac5e2f82
674129,1,43d8ac5e2f82
107517,1,43d8ac5e2f82
607647,1,43d8ac5e2f82
34446,1,43d8ac5e2f82
158934,1,43d8ac5e2f82
378126,1,43d8ac5e2f82
8346,1,43d8ac5e2f82
800249,1,43d8ac5e2f82
569598,1,43d8ac5e2f82
410,1,43d8ac5e2f82
5608,1,43d8ac5e2f82
482187,1,43d8ac5e2f82
640988,1,43d8ac5e2f82
675155,1,43d8ac5e2f82
801456,1,43d8ac5e2f82
10028,1,43d8ac5e2f82
496397,1,43d8ac5e2f82
9168,1,43d8ac5e2f82
557128,1,43d8ac5e2f82
15987,1,43d8ac5e2f82
552001,1,43d8ac5e2f82
822177,1,43d8ac5e2f82
831,1,43d8ac5e2f82
837840,1,43d8ac5e2f82
513013,1,43d8ac5e2f82
378124,1,43d8ac5e2f82
680735,1,43d8ac5e2f82
336110,1,43d8ac5e2f82
5033,1,43d8ac5e2f82
12678,1,43d8ac5e2f82
282932,1,43d8ac5e2f82
838486,1,43d8ac5e2f82
512463,1,43d8ac5e2f82
136195,1,43d8ac5e2f82
822002,1,43d8ac5e2f82
175247,1,43d8ac5e2f82
615893,1,43d8ac5e2f82
387813,1,43d8ac5e2f82
5092,1,43d8ac5e2f82
4928,1,43d8ac5e2f82
16700,1,43d8ac5e2f82
96982,1,43d8ac5e2f82
589226,1,43d8ac5e2f82
14841,1,43d8ac5e2f82
648866,1,43d8ac5e2f82
12627,1,43d8ac5e2f82
232020,1,43d8ac5e2f82
21879,1,43d8ac5e2f82
325008,1,43d8ac5e2f82
262187,1,43d8ac5e2f82
377553,1,43d8ac5e2f82
377381,1,43d8ac5e2f82
650189,1,43d8ac5e2f82
17500,1,43d8ac5e2f82
832482,1,43d8ac5e2f82
800040,1,43d8ac5e2f82
16051,1,43d8ac5e2f82
663652,1,43d8ac5e2f82
3736,1,43d8ac5e2f82
313,1,43d8ac5e2f82
612236,1,43d8ac5e2f82
15323,1,43d8ac5e2f82
17322,1,43d8ac5e2f82
498334,1,43d8ac5e2f82
640987,1,43d8ac5e2f82
676071,1,43d8ac5e2f82
12908,1,43d8ac5e2f82
10373,1,43d8ac5e2f82
850882,1,43d8ac5e2f82
351780,1,43d8ac5e2f82
35136,1,43d8ac5e2f82
838696,1,43d8ac5e2f82
176329,1,43d8ac5e2f82
822101,1,43d8ac5e2f82
376052,1,43d8ac5e2f82
845584,1,43d8ac5e2f82
130284,1,43d8ac5e2f82
845579,1,43d8ac5e2f82
328378,1,43d8ac5e2f82
699490,1,43d8ac5e2f82
354,1,43d8ac5e2f82
501336,1,43d8ac5e2f82
18713,1,43d8ac5e2f82
518514,1,43d8ac5e2f82
15878,1,43d8ac5e2f82
572953,1,43d8ac5e2f82
61105,1,43d8ac5e2f82
319547,1,43d8ac5e2f82
10867,1,43d8ac5e2f82
8562,1,43d8ac5e2f82
228771,1,43d8ac5e2f82
10388,1,43d8ac5e2f82
\end{filecontents*}
\csvreader[late after line=\\\hline, late after last line=\\, head to column names, filter={\value{csvrow}<5}]
            {anc/stamps-fm-3-dataProdutoComercianteQuota.csv}
            {}
            {\codigo & \quota & \comerciante}
        \bottomrule
    \end{tabular}
    \caption{Sample of the list of products by merchant}
    \label{tab:products}
\normalsize\end{table}


In \cref{tab:influencers,tab:merchants,tab:products}, we present a sample where we can observe that each influencer can represent one or more products, and each merchant can accommodate up to a certain number of influencers and offer their products, which in turn can accommodate a specific number of influencers simultaneously. The total number of desired influencers may be less than, equal to, or greater than the supported number, but we stipulate $\max \{ l_j: p_j \in P_k \} \le q_k \le \sum \{ l_j: p_j \in P_k \}$ as the operational limit.


An allocation $E$ is a subset of $I \times P$ such that $(f_i, p_j) \in E \Longrightarrow p_j \in D_i$, meaning that $f_i$ wants to advertise $p_j$, and for each influencer $f_i \in F$, $\mid \{ ( f_i, p_j ) \in E: p_j \in P \} \mid \le 1$. Influencer $f_i$ has been recommended product $p_j$, and product $p_j$ has been recommended to influencer $f_i$ if $(f_i, p_j) \in E$. We can also say that if $f_i$ has been recommended $p_j$ in $E$, where $p_j \in P_k$, then $f_i$ has been recommended to $v_k$, and $v_k$ has been recommended to $f_i$.


If an influencer $f_i \in F$, $f_i$ has been paired in $E$ with some product $p_j$, then $E(f_i)$ represents $p_j$, or conversely, $f_i$ does not exist in $E$.


For a given product, the expression $p_j \in P$, $E(p_j)$ represents the set of influencers allocated with $p_j$ in $E$. Product $p_j$ can be under-allocated, fully allocated, or over-allocated if $\mid E(p_j) \mid$ is respectively less than, equal to, or greater than $l_j$. Similarly, for any merchant $v_k \in V$, $\mid E(v_k) \mid$ represents the set of influencers recommended to $v_k$ in $E$. Merchant $v_k$ is under-allocated, fully allocated, or over-allocated if $\mid E(v_k) \mid$ is respectively less than, equal to, or greater than $q_k$.


A matching $E$ is an allocation such that the allocation of each product satisfies the condition $p_j \in P, \mid E(p_j) \le l_j$, and the allocation of each merchant satisfies the condition $v_k \in V, \mid E(v_k) \mid \le q_k$.


Thus, $p_j$ is allocated to a maximum of $l_j$ influencers in $E$ as long as it does not exceed $q_k$ for $v_k$ in $E$.


The pair $(f_i, p_j) \in (F \times P)$ is subject to $p_j \in D_i$ (influencer $f_i$ desires $p_j$) and $f_i \notin E$ or $f_i$ prefers $p_j$ over $E(f_i)$.


Both $p_j$ and $v_k$ are under-allocated, or $p_j$ is under-allocated and $v_k$ is fully allocated, and either $f_i \in E(v_k)$ or $v_k$ prefers $f_i$ to the least qualified influencer in $E(v_k)$, or $p_j$ is fully allocated and $v_k$ prefers $f_i$ to the least qualified influencer in $E(p_j)$ - we say this case of $(f_i, p_j)$ is a blocking pair in $E$. A matching is stable if there are no blocking pairs.

\section*{Proof of Concept}

\subsection*{Approach Description}\label{subsec:approach}


To extract lists of potential influencers, products, and merchants available in the dataset, performance indicators were simulated as a way to demonstrate an application of the solution:

\begin{enumerate}
    \item Influencers $f_i$, reputation $r(f_i)$, and list of desired products $E$ based on different prioritization criteria:
        \begin{enumerate}
            \item Reputation $r(f_i)$ from the \gls{GT}, i.e., the more the consumer spends on a particular product, the higher their reputation for that product (see \cref{tab:influencers})
            \item Reputation $r(f_i)$ from the \gls{FMC} (see \cref{tab:influencersmc}), given by the formula:
                \[
                    r(f_i) = K / U,
                \]
                where:
                \begin{enumerate}
                    \item $K$: Number of orders
                    \item $U$: Total consumers or potential influencers
                \end{enumerate}
        \end{enumerate}
    \item Merchants $v_k$ and their respective quotas $q_k$:
        \begin{enumerate} 
            \item ``Quota'' $q_k$ calculated based on the total volume sold, i.e., the more the merchant $v_k$ sells, the more space for influencers they will have (see \cref{tab:merchants})
        \end{enumerate}
    \item Products $p_j$, quota $l_j$ per product and merchant $v_k$:
        \begin{enumerate} 
            \item Quota $l_j$ for product $p_j$ assigned manually, i.e., based on the total quota $l_j$ of merchant $v_k$, they can request influencers for their products (see \cref{tab:products})
        \end{enumerate}
\end{enumerate}

\begin{quote}
Note that the value assigned to $r(f_i)$ is not important - the descending order of this value is sufficient for the algorithm to prioritize an influencer.
\end{quote}

\subsection*{Experiments}

To validate this proposal, we collected data from different sources:


\begin{enumerate}
    \item Public datasets from the internet\footnote{See \url{https://www.kaggle.com}}
        \begin{enumerate}
            \item Electronics sales\cite{mkechinov-ecommerce-purchase}
            \item Supermarket sales\cite{aungpyaeap-supermarket-sales}
        \end{enumerate}
\end{enumerate}


The importance of sorting the data lies in the fact that, in general, the "Gale-Shapley" algorithm terminates in most cases in $n^2$ iterations, with a worst-case computational time of $\Omega(n^2)$\cite{kleinberg2006algorithm}. Thus, it is desirable to reduce the number of elements analyzed to consider only the most interesting ones. For this purpose, we use criteria based on performance indicators (see \nameref{subsec:approach}) aligned with organizational objectives.


In the supermarket database, there is no identification of the buyer (i.e., potential influencer). In this situation, a profile was defined composed of the consumer's information such as \textit{branch}, \textit{city}, \textit{type}, and \textit{gender}.

\subsection*{Results}

\subsubsection*{Kaggle Dataset - Electronics Store - Average Purchase Frequency X Total Spending}

\begin{figure*}[ht]
    \noindent\begin{minipage}[c]{0.5\textwidth}
        \centering
        \includegraphics[width=.9\textwidth]{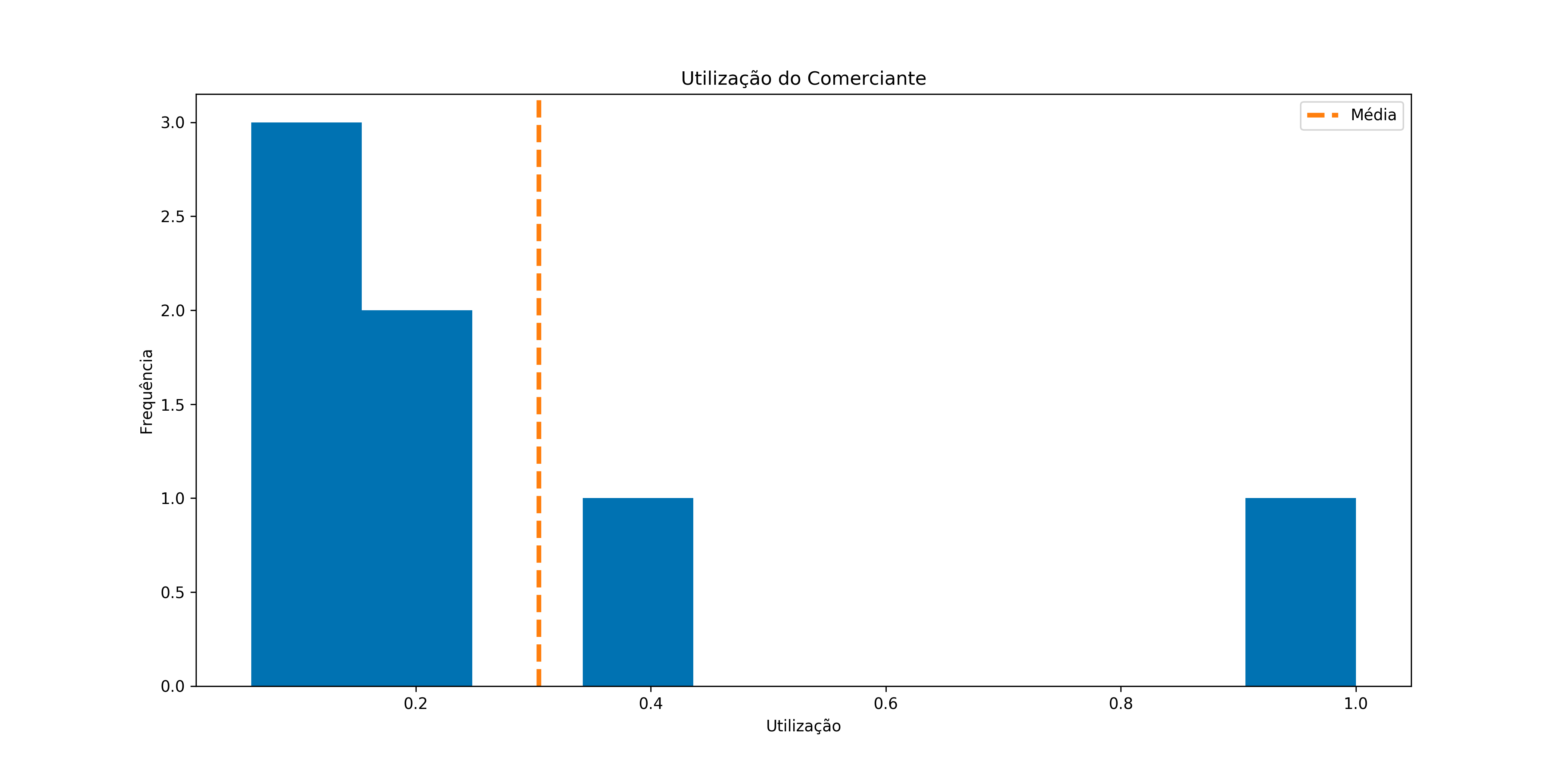}
        \caption*{by \gls{FMC}}
    \end{minipage}
    \begin{minipage}[c]{0.5\textwidth}
        \centering
        \includegraphics[width=.9\textwidth]{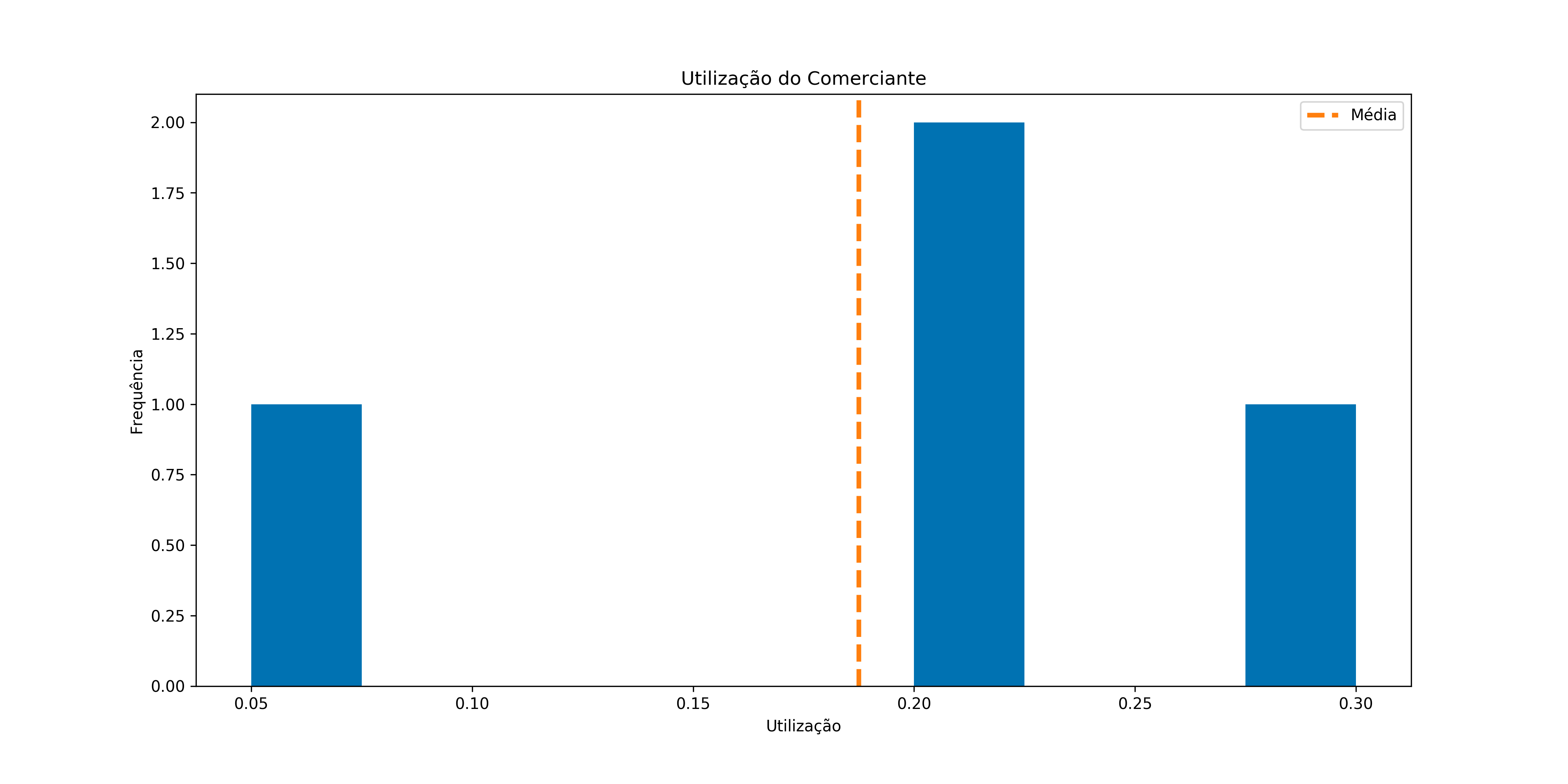}
        \caption*{by \gls{GT}}
    \end{minipage}
    \caption{Merchant Utilization}\label{fig:img-eletr-utilizacao-comerciante}
\end{figure*}

Using data from the electronics store\cite{mkechinov-ecommerce-purchase}, we observe in \cref{fig:img-eletr-utilizacao-comerciante} the low efficiency in recommending influencers to merchants. This result is a consequence of the fact that electronic products are not repeatedly purchased by the same consumer.

\begin{figure*}[ht]
    \noindent\begin{minipage}[c]{0.5\textwidth}
        \centering
        \includegraphics[width=.9\textwidth]{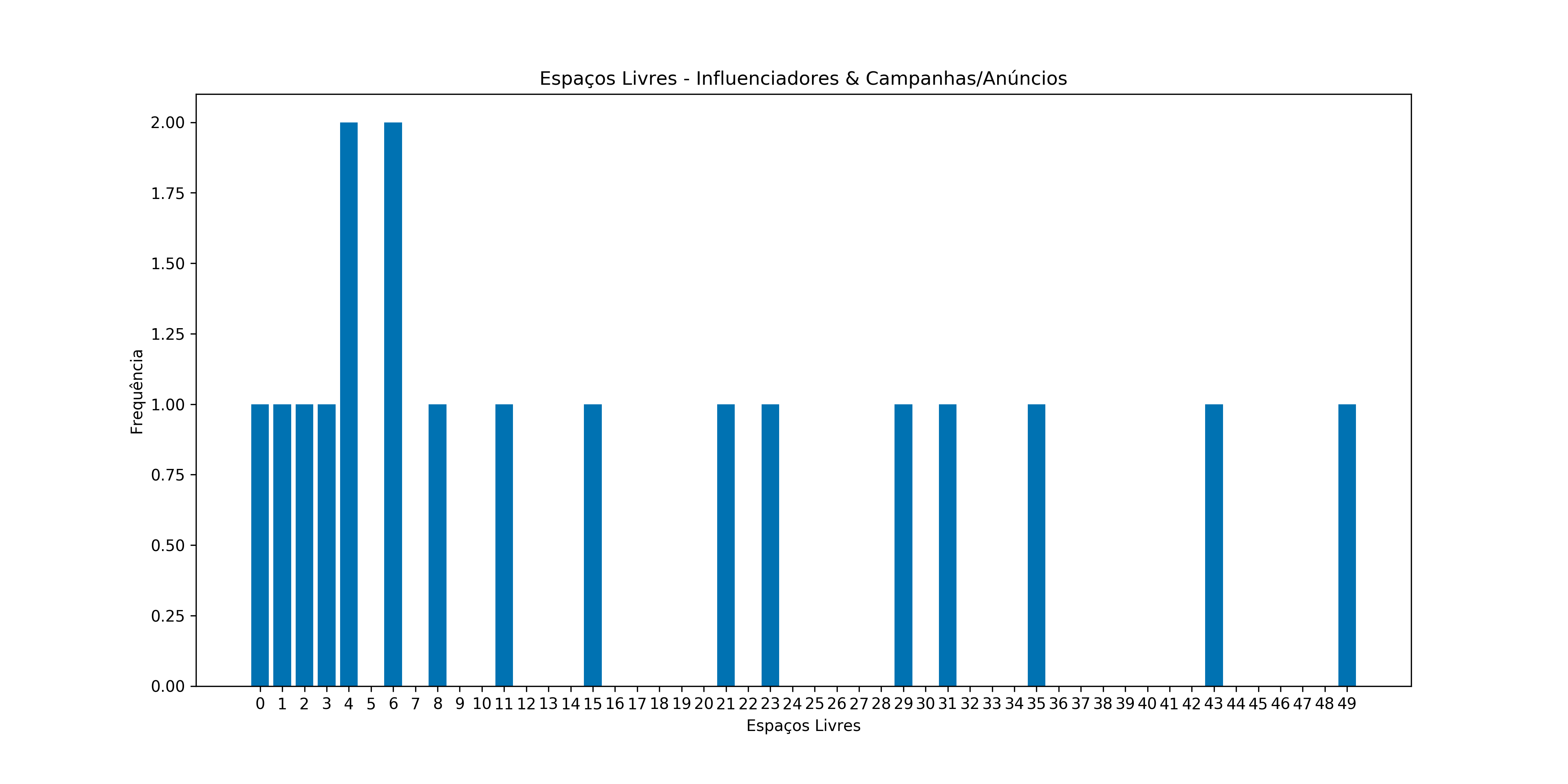}
        \caption*{by \gls{FMC}}
    \end{minipage}
    \begin{minipage}[c]{0.5\textwidth}
        \centering
        \includegraphics[width=.9\textwidth]{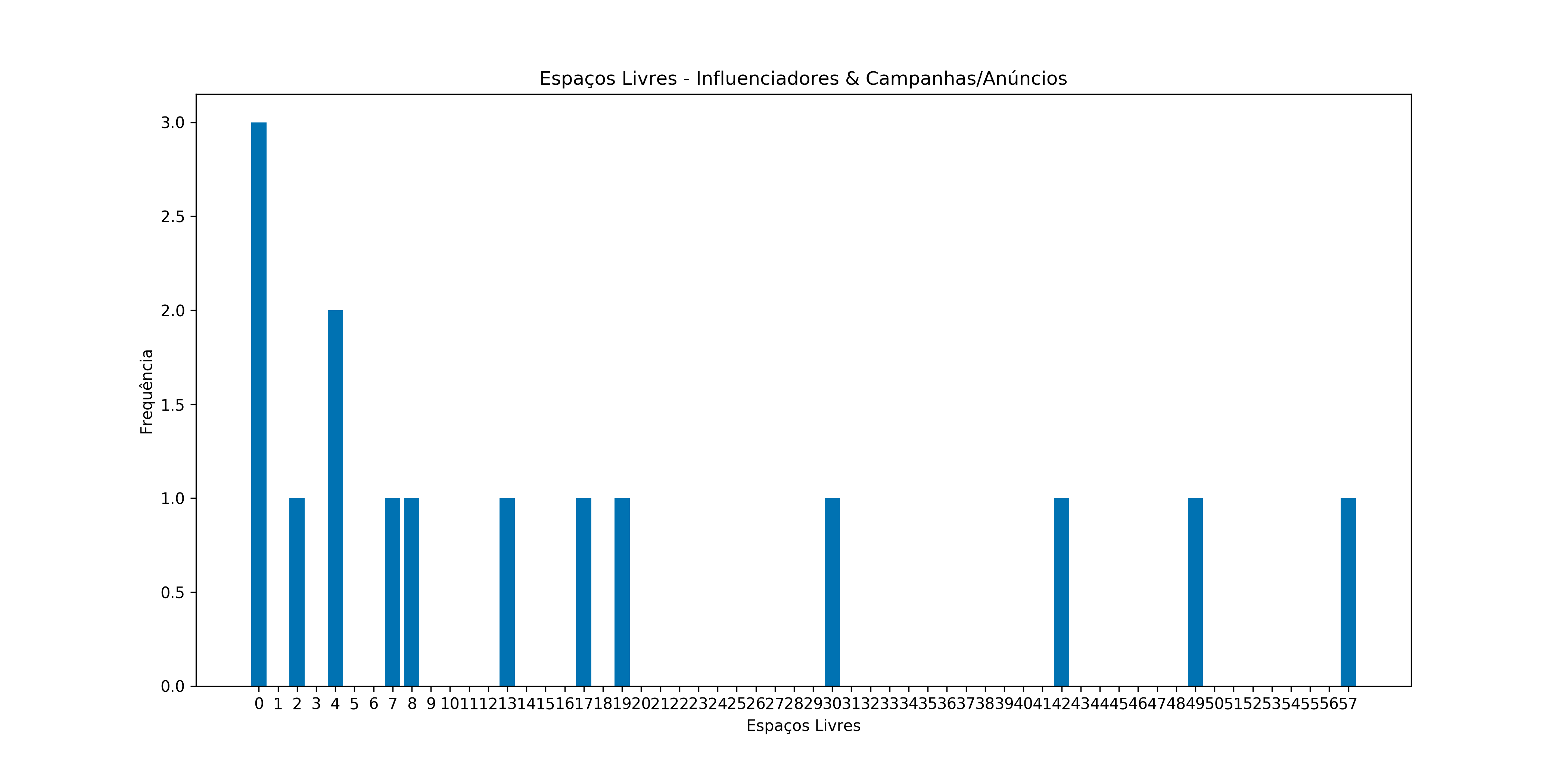}
        \caption*{by \gls{GT}}
    \end{minipage}
    \caption{Free slots}\label{fig:img-eletr-anuncios-espacos}
\end{figure*}

Given the sales volume of some manufacturers (see \cref{tab:merchants}), the maximum quota of influencers ends up being high due to $\sum \{ l_j: p_j \in P_k \}$ (see \nameref{subsec:definicao}). This is reflected in the high frequency of empty spaces that we can observe in \cref{fig:img-eletr-anuncios-espacos}.

\begin{figure*}[ht]
    \noindent\begin{minipage}[c]{0.5\textwidth}
        \centering
        \includegraphics[width=.9\textwidth]{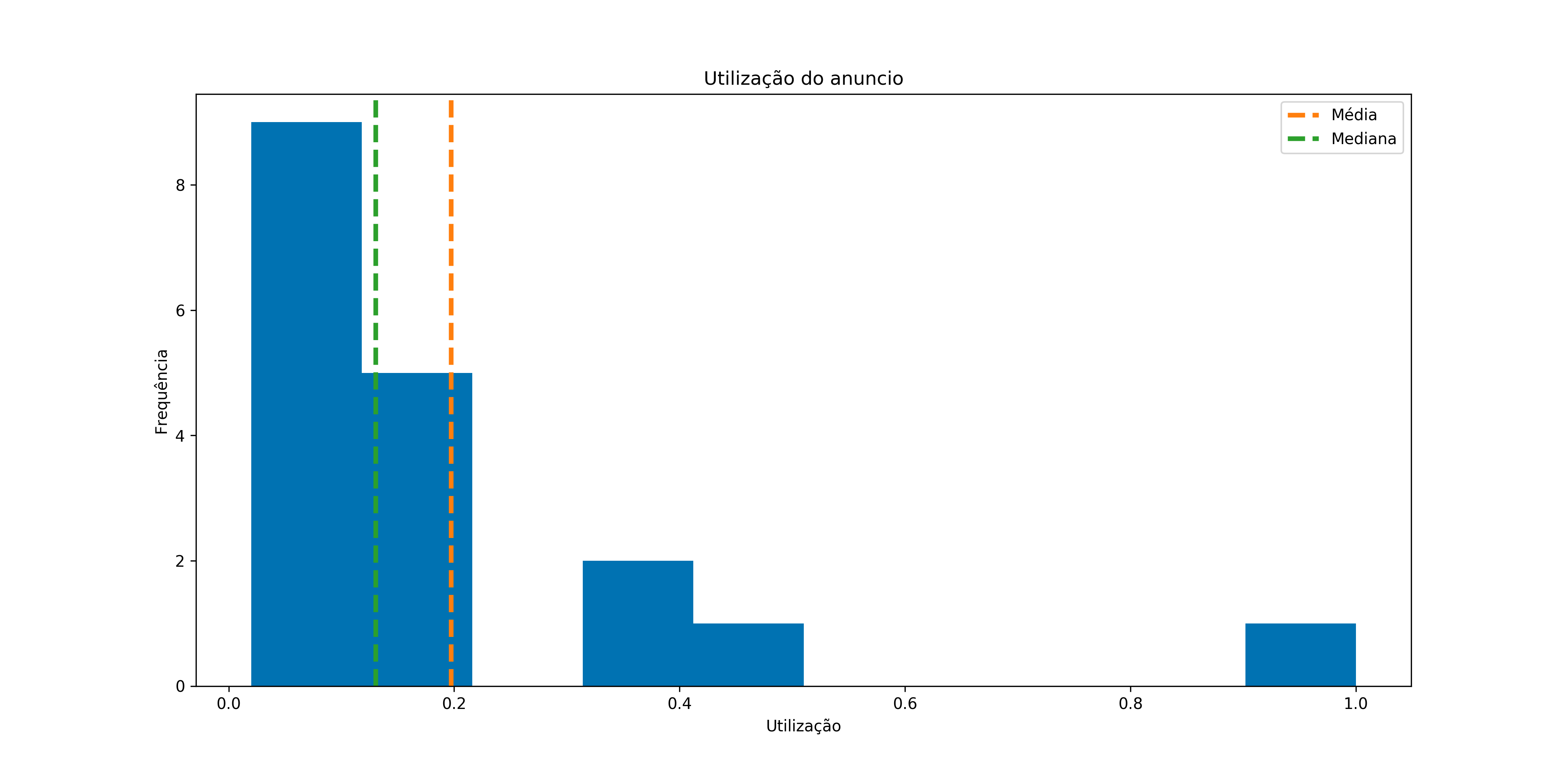}
        \caption*{by \gls{FMC}}
    \end{minipage}
    \begin{minipage}[c]{0.5\textwidth}
        \centering
        \includegraphics[width=.9\textwidth]{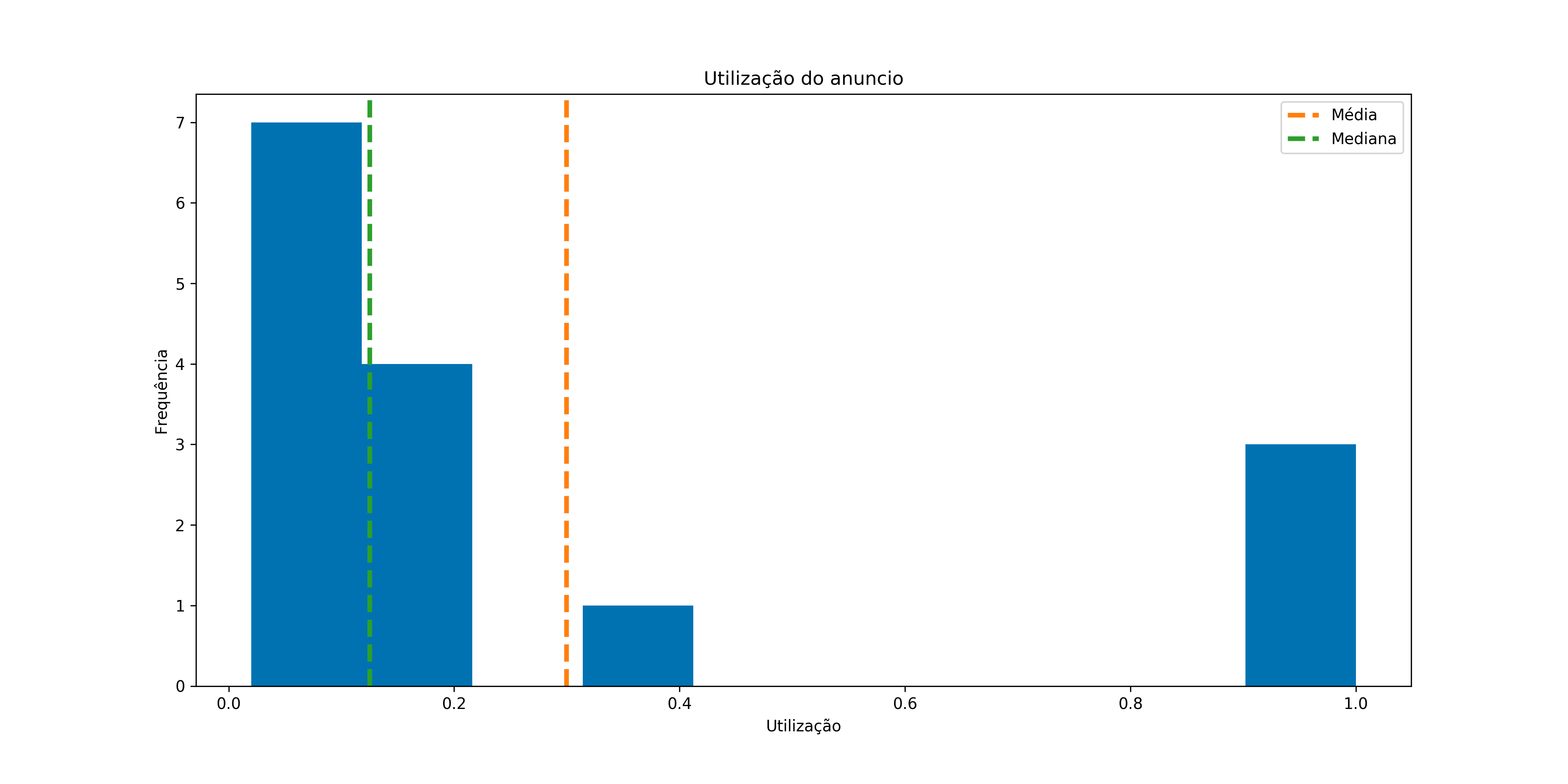}
        \caption*{by \gls{GT}}
    \end{minipage}
    \caption{Advertisement Usage}\label{fig:img-eletr-anuncios-utilizacao}
\end{figure*}

In \cref{fig:img-eletr-anuncios-utilizacao}, we see a concentration in the low utilization of advertisements. The wide variety of different products does not allow for specific concentration.

\begin{figure*}[ht]
    \noindent\begin{minipage}[c]{0.5\textwidth}
        \centering
        \includegraphics[width=.9\textwidth]{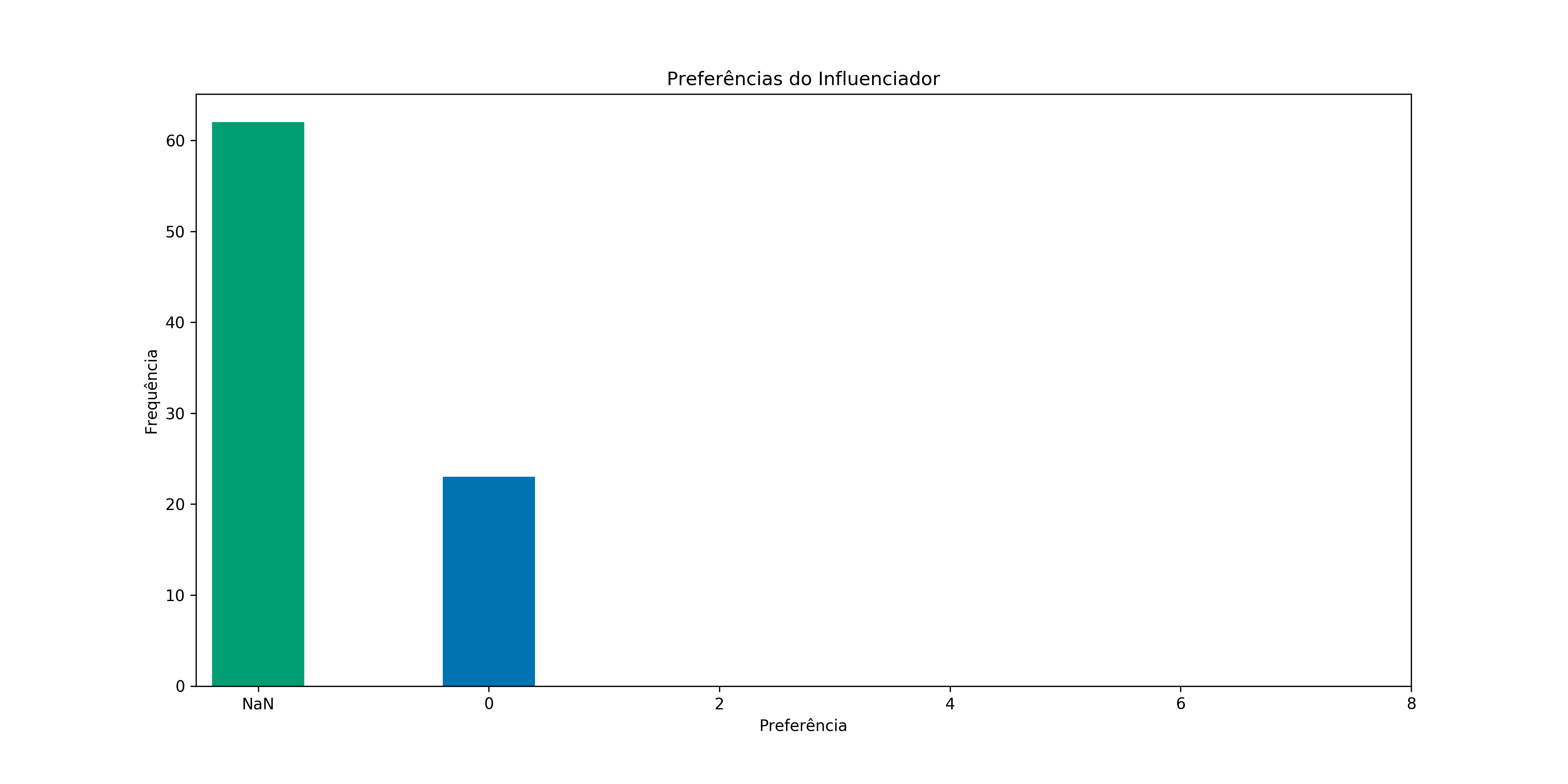}
        \caption*{por \gls{FMC}}
    \end{minipage}
    \begin{minipage}[c]{0.5\textwidth}
        \centering
        \includegraphics[width=.9\textwidth]{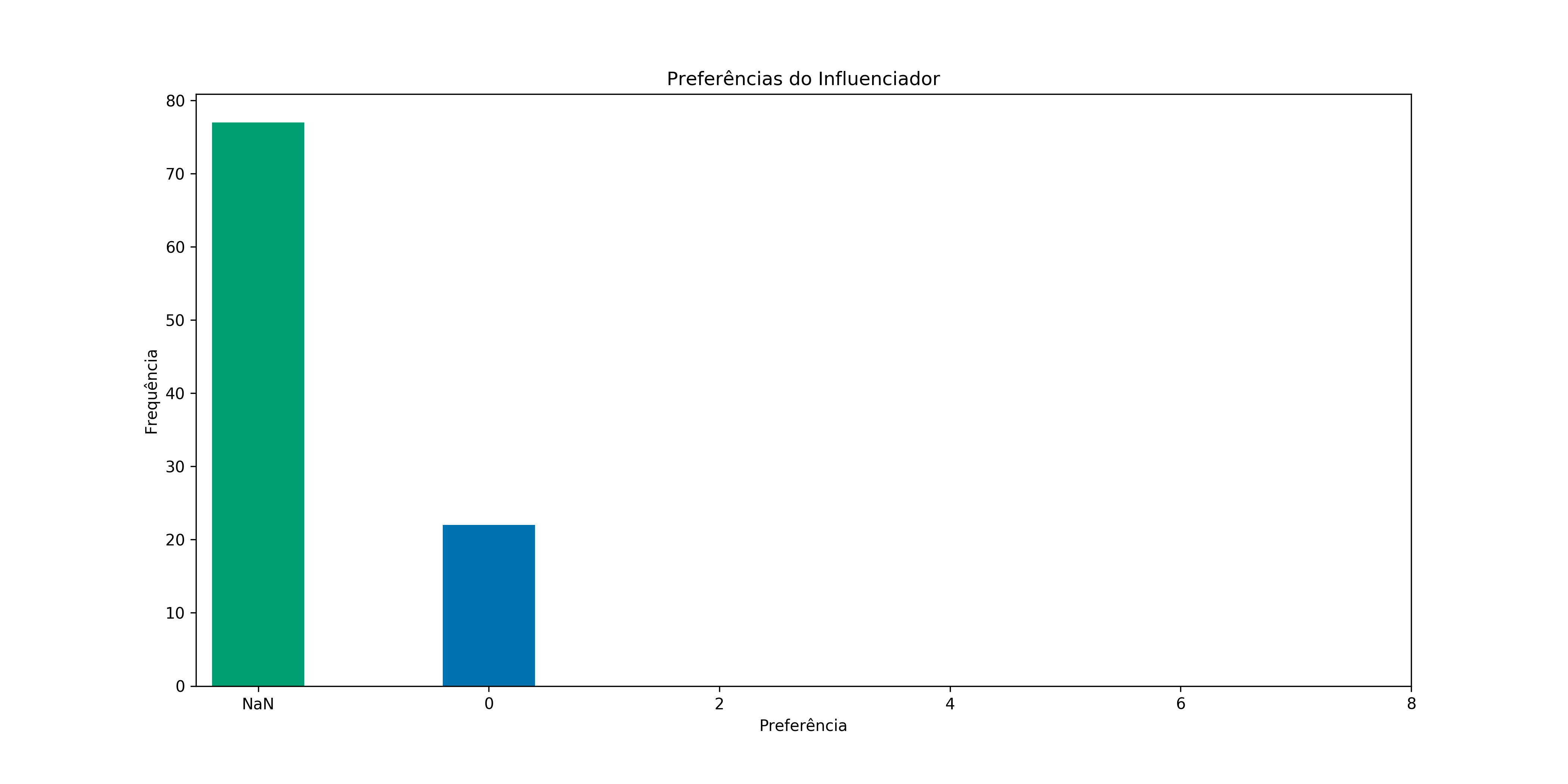}
        \caption*{por \gls{GT}}
    \end{minipage}
    \caption{Preferência do influenciador}\label{fig:img-eletr-influenciador-preferencias}
\end{figure*}

With the assignment of a preference for consumers and as a consequence of the great diversity of different products, it is noted that the assignment is not able to indicate consumer preference (as can be seen in \cref{fig:img-eletr-influenciador-preferencias}).

\begin{figure*}[ht]
    \noindent\begin{minipage}[c]{0.5\textwidth}
        \centering
        \includegraphics[width=.9\textwidth]{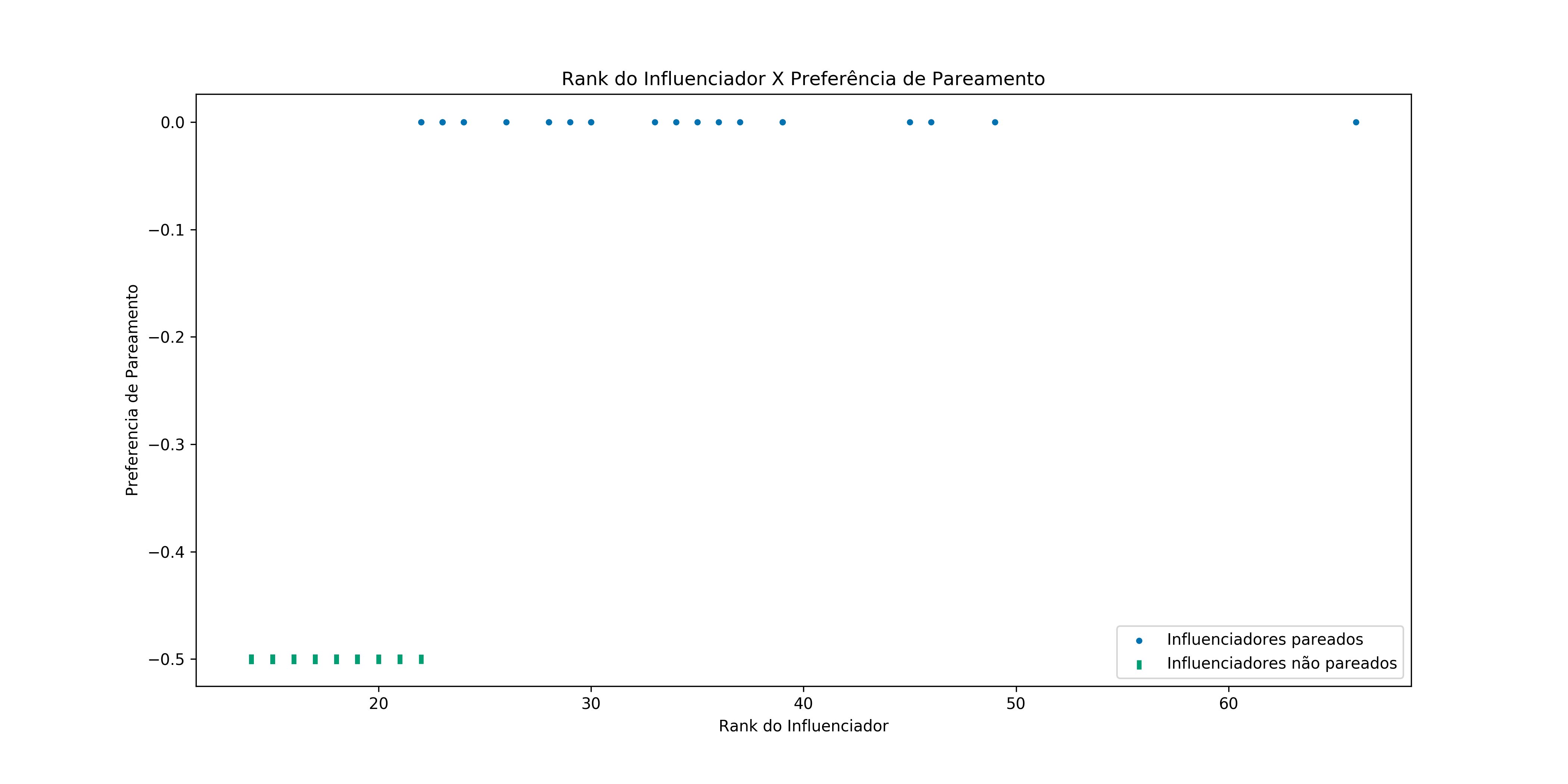}
        \caption*{por \gls{FMC}}
    \end{minipage}
    \begin{minipage}[c]{0.5\textwidth}
        \centering
        \includegraphics[width=.9\textwidth]{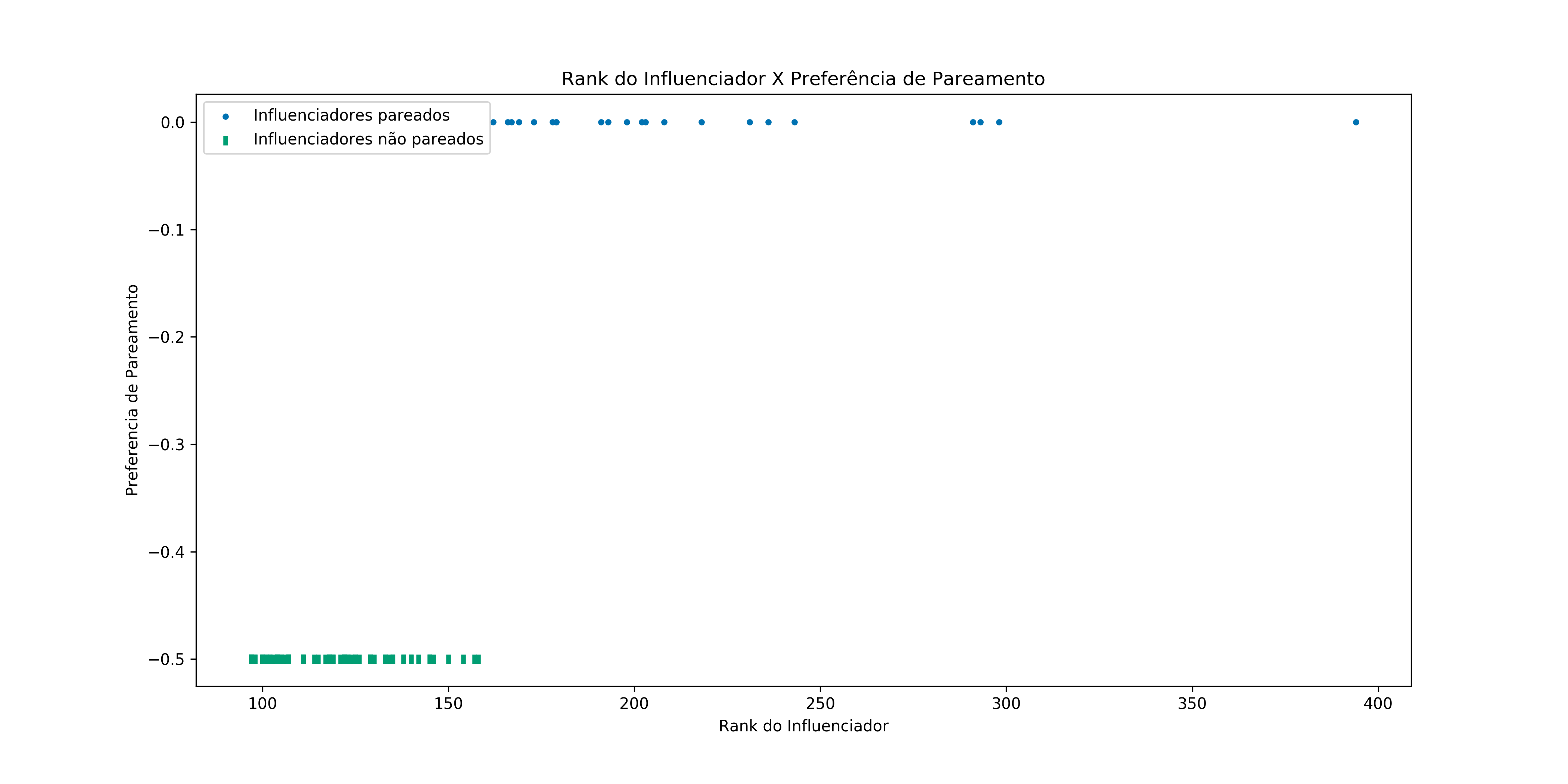}
        \caption*{por \gls{GT}}
    \end{minipage}
    \caption{Rank influenciador e preferência}\label{fig:img-eletr-influenciador-preferencia-rank}
\end{figure*}

The consumer classification by \gls{FMC} or \gls{GT} is unable to bring a significant change in the assignment of their preference in indicating them as a potential influencer for a particular product (see in \cref{fig:img-eletr-influenciador-preferencia-rank}).

\subsubsection*{Kaggle Dataset - Supermarket - Average Purchase Frequency vs Total Spending}

\begin{figure*}[ht]
    \noindent\begin{minipage}[c]{0.5\textwidth}
        \centering
        \includegraphics[width=.9\textwidth]{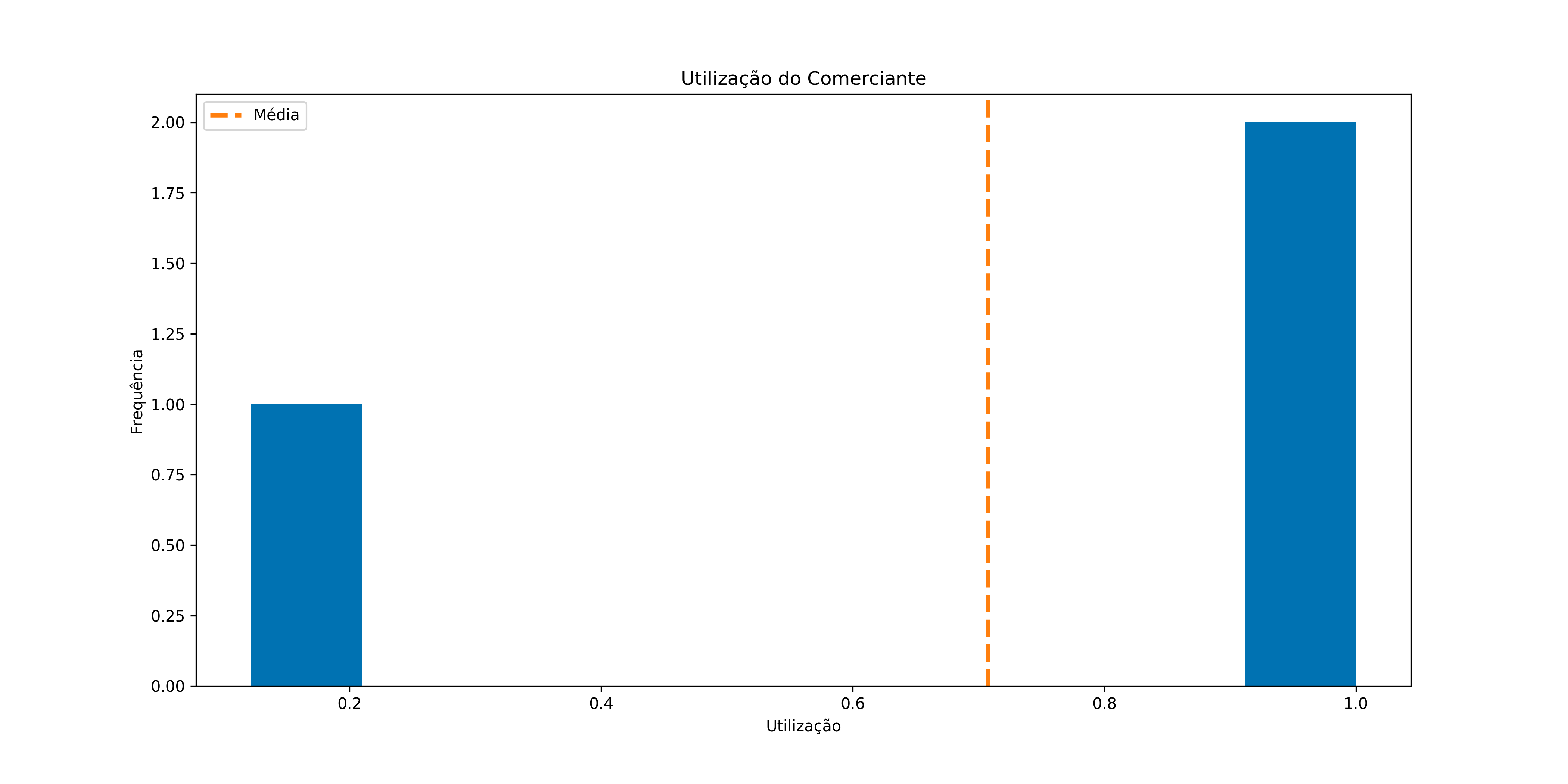}
        \caption*{by \gls{FMC}}
    \end{minipage}
    \begin{minipage}[c]{0.5\textwidth}
        \centering
        \includegraphics[width=.9\textwidth]{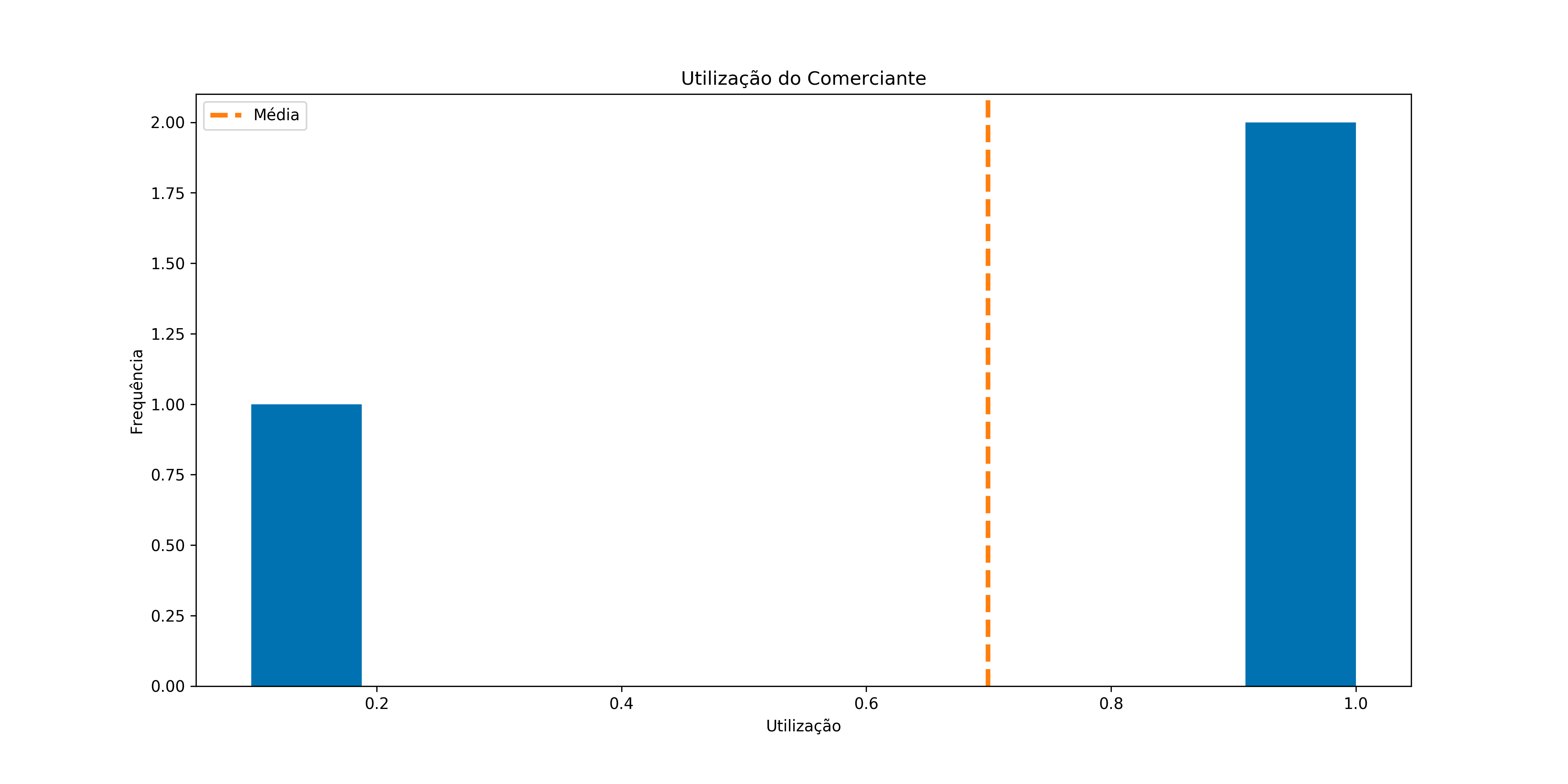}
        \caption*{by \gls{GT}}
    \end{minipage}
    \caption{Merchant Utilization}\label{fig:img-super-utilizacao-comerciante}
\end{figure*}

Using supermarket sales data\cite{aungpyaeap-supermarket-sales}, we can see in \cref{fig:img-super-utilizacao-comerciante} the separation between unpopular products on the left with low ad recommendation frequency and popular products on the right.

\begin{figure*}[ht]
    \noindent\begin{minipage}[c]{0.5\textwidth}
        \centering
        \includegraphics[width=.9\textwidth]{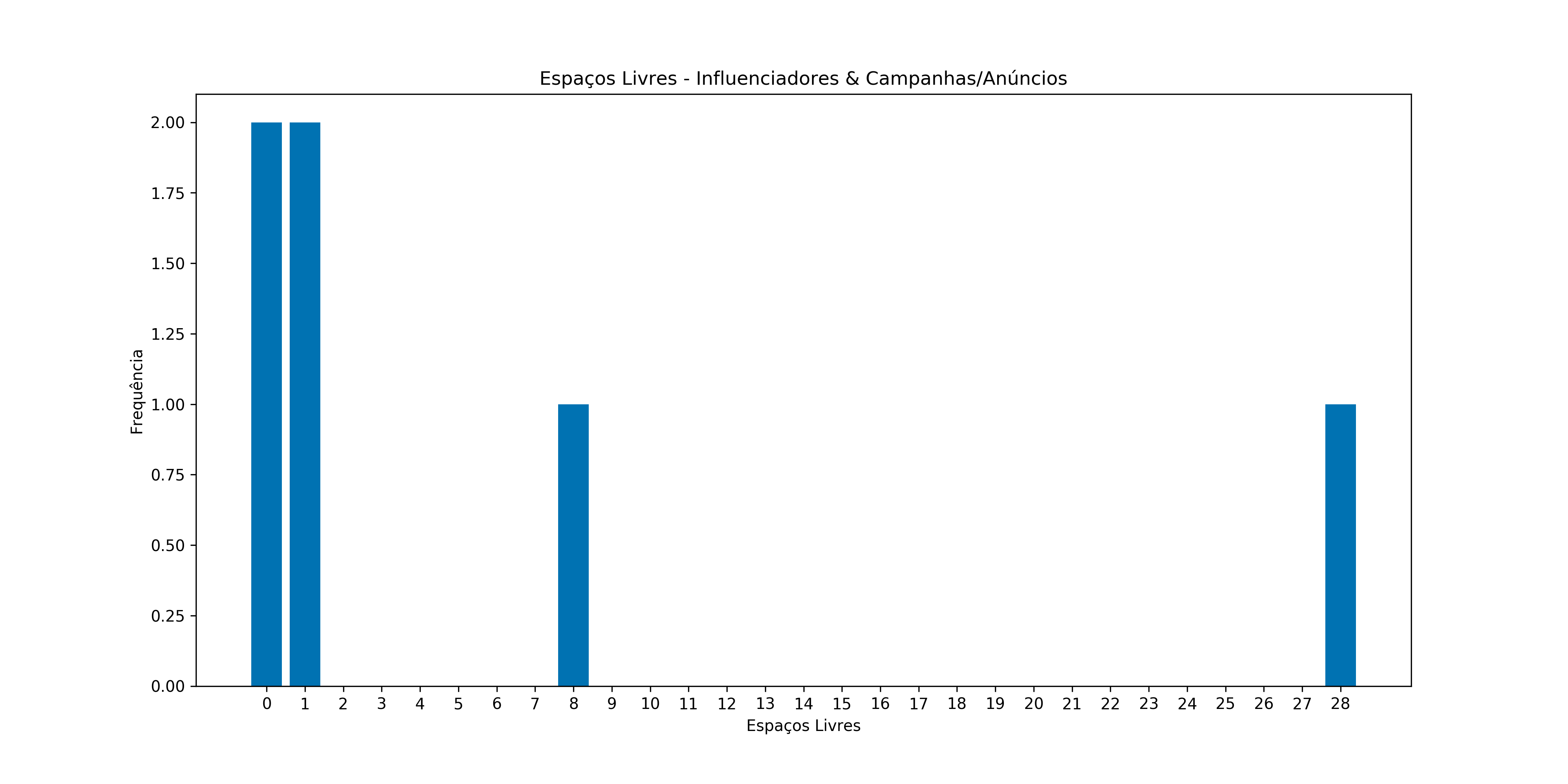}
        \caption*{by \gls{FMC}}
    \end{minipage}
    \begin{minipage}[c]{0.5\textwidth}
        \centering
        \includegraphics[width=.9\textwidth]{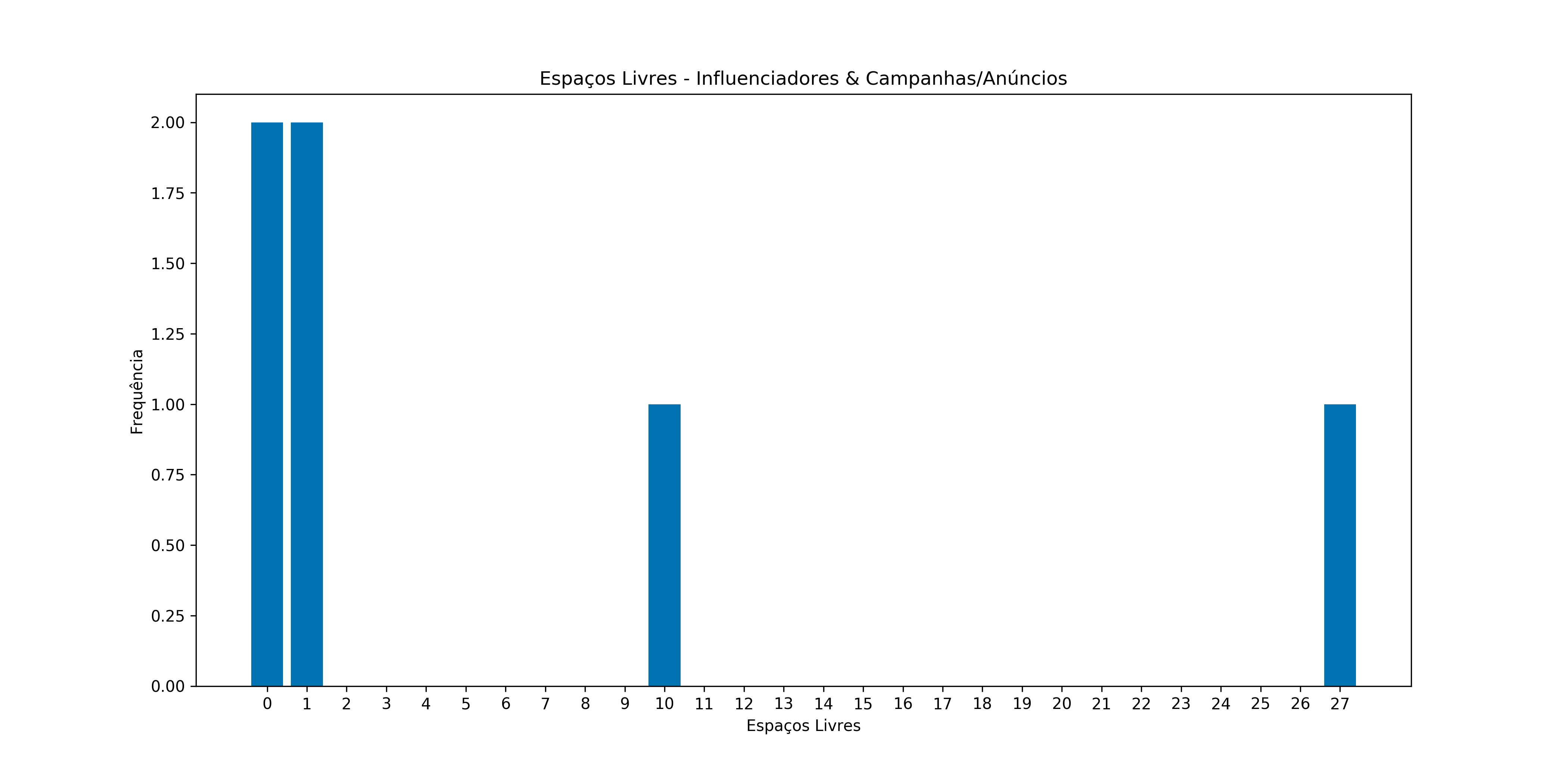}
        \caption*{by \gls{GT}}
    \end{minipage}
    \caption{Free slots}\label{fig:img-super-anuncios-espacos}
\end{figure*}

The initial recommendations for the consumer are mostly satisfied, with some extreme cases of many empty spaces, as observed in \cref{fig:img-super-anuncios-espacos}.

\begin{figure*}[ht]
    \noindent\begin{minipage}[c]{0.5\textwidth}
        \centering
        \includegraphics[width=.9\textwidth]{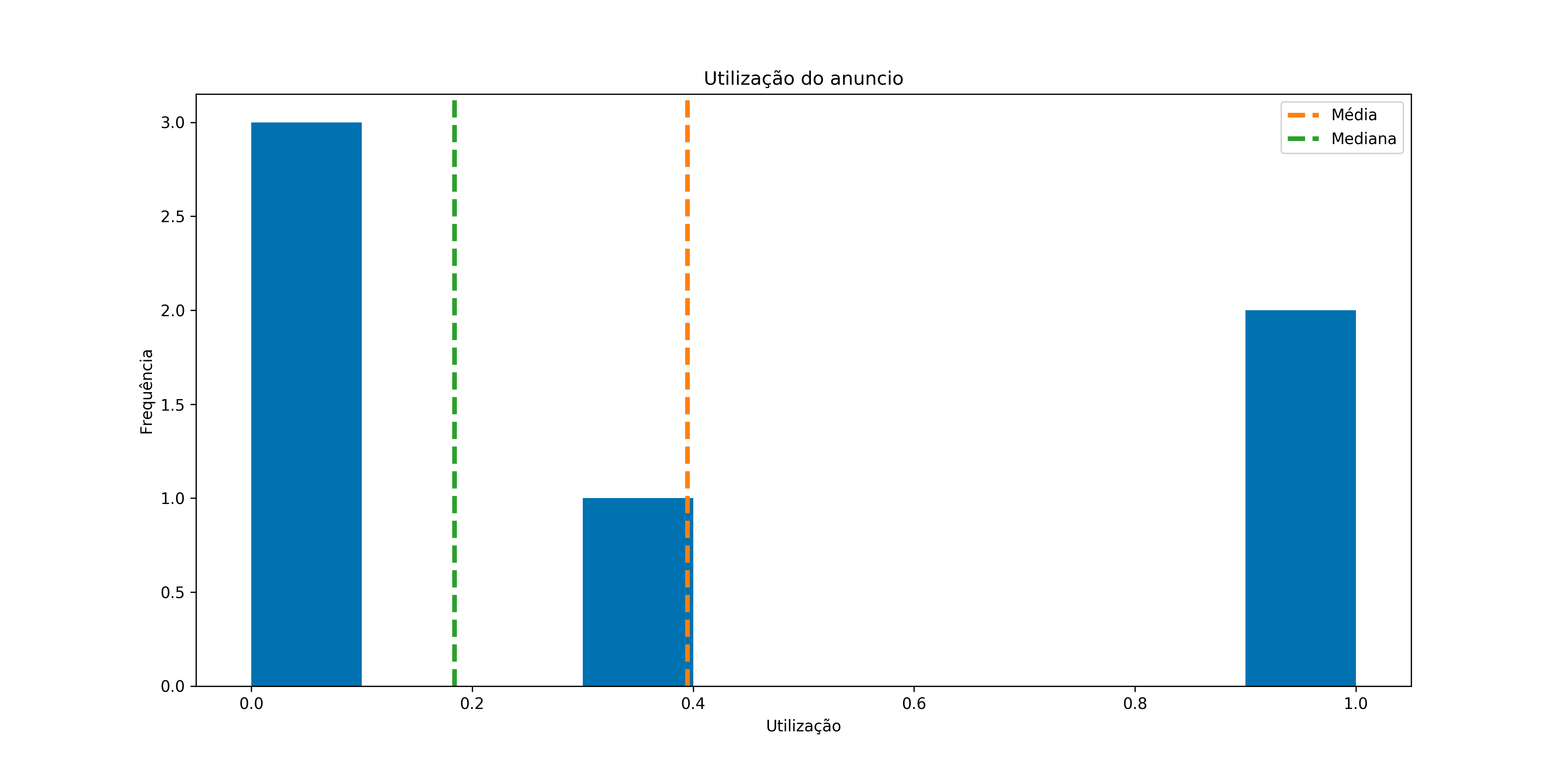}
        \caption*{by \gls{FMC}}
    \end{minipage}
    \begin{minipage}[c]{0.5\textwidth}
        \centering
        \includegraphics[width=.9\textwidth]{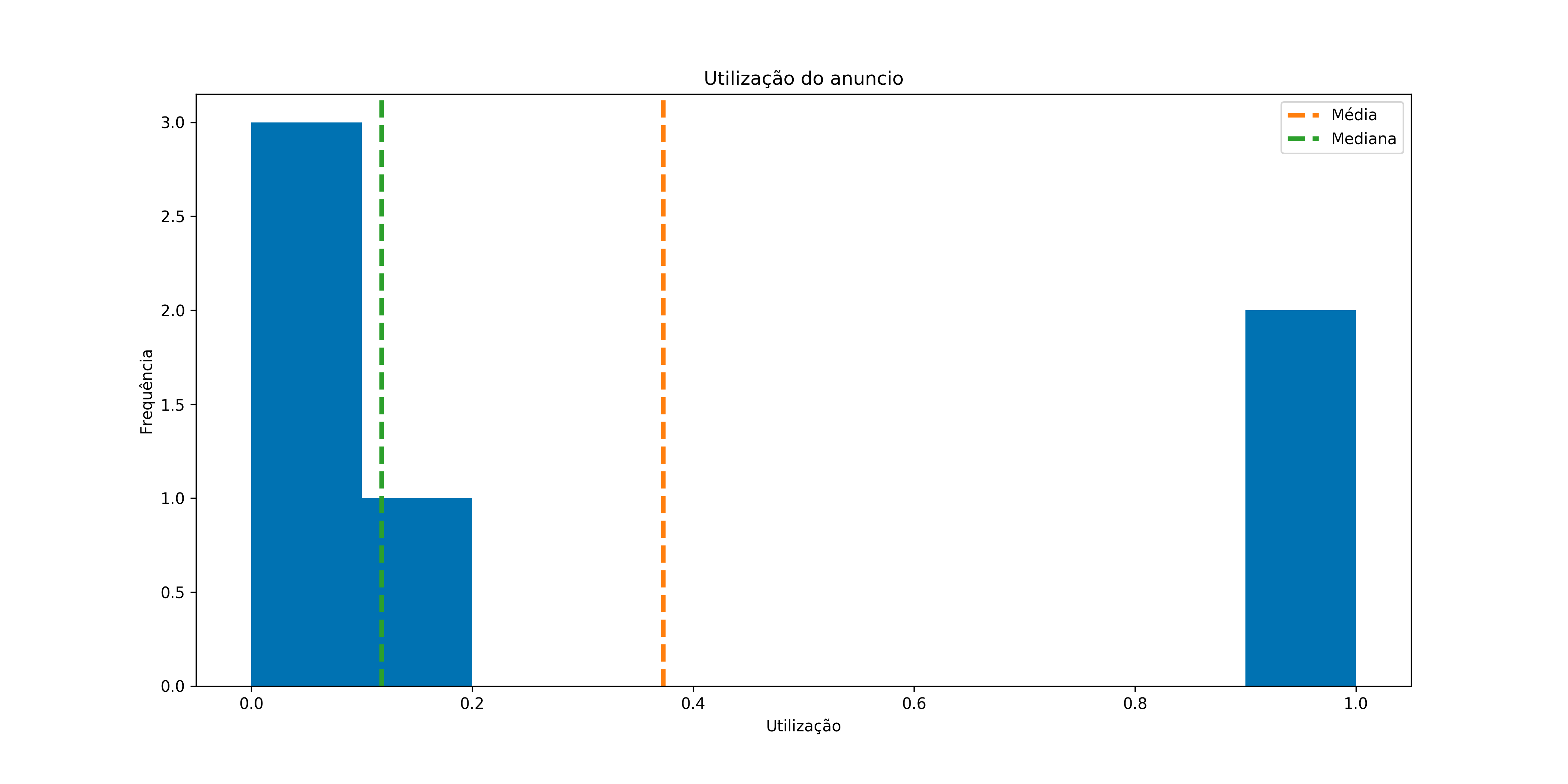}
        \caption*{by \gls{GT}}
    \end{minipage}
    \caption{Ad Utilization}\label{fig:img-super-anuncios-utilizacao}
\end{figure*}

In \cref{fig:img-super-anuncios-utilizacao}, we observe the distribution on the left of unpopular products and on the right of popular products in the recommendations for consumers.

\begin{figure*}[ht]
    \noindent\begin{minipage}[c]{0.5\textwidth}
        \centering
        \includegraphics[width=.9\textwidth]{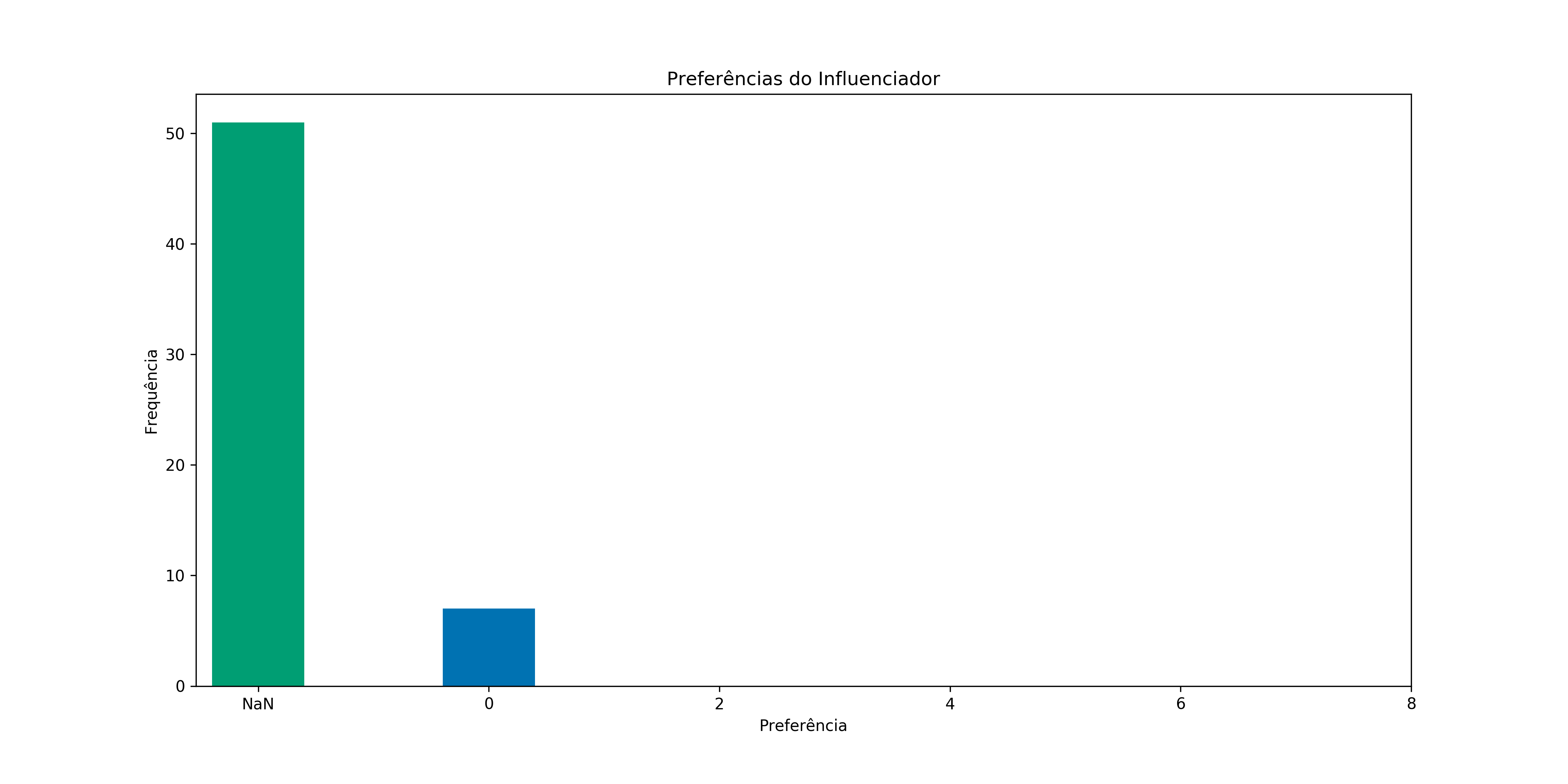}
        \caption*{by \gls{FMC}}
    \end{minipage}
    \begin{minipage}[c]{0.5\textwidth}
        \centering
        \includegraphics[width=.9\textwidth]{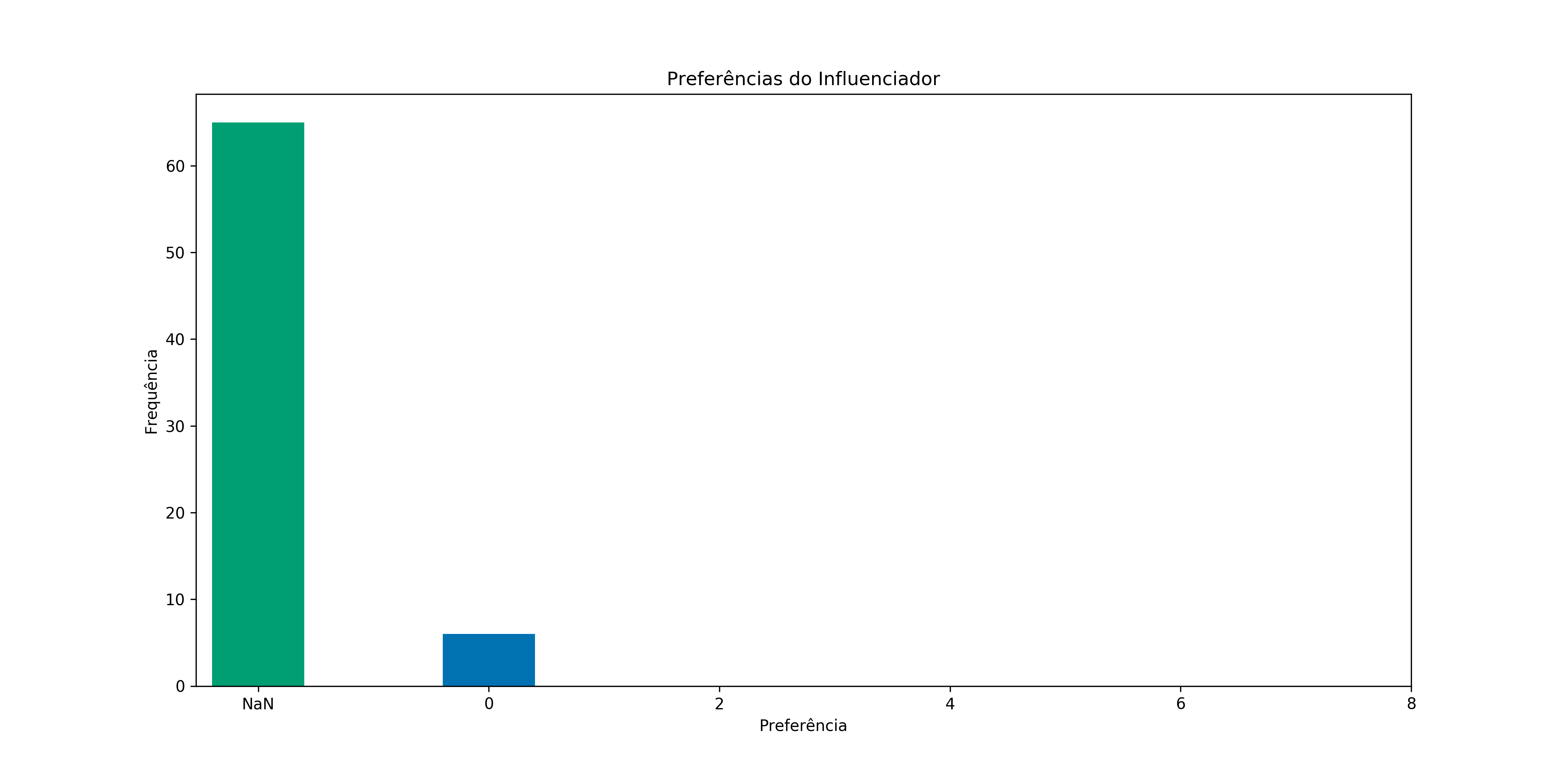}
        \caption*{by \gls{GT}}
    \end{minipage}
    \caption{Influencer Preference}\label{fig:img-super-influenciador-preferencias}
\end{figure*}

We can see in \cref{fig:img-super-influenciador-preferencias} that there was no clear indication of consumer preference.

\begin{figure*}[ht]
    \noindent\begin{minipage}[c]{0.5\textwidth}
        \centering
        \includegraphics[width=.9\textwidth]{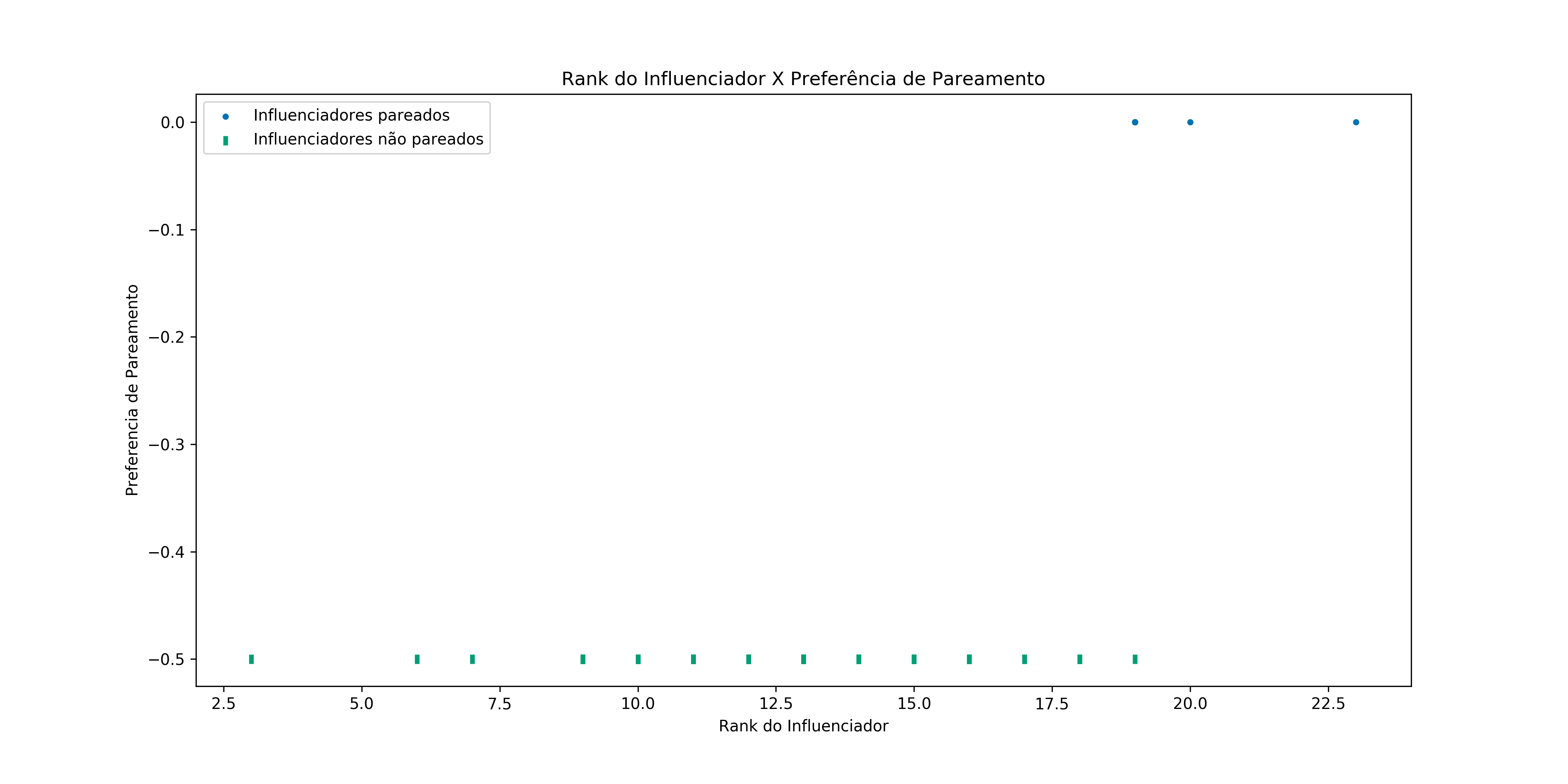}
        \caption*{by \gls{FMC}}
    \end{minipage}
    \begin{minipage}[c]{0.5\textwidth}
        \centering
        \includegraphics[width=.9\textwidth]{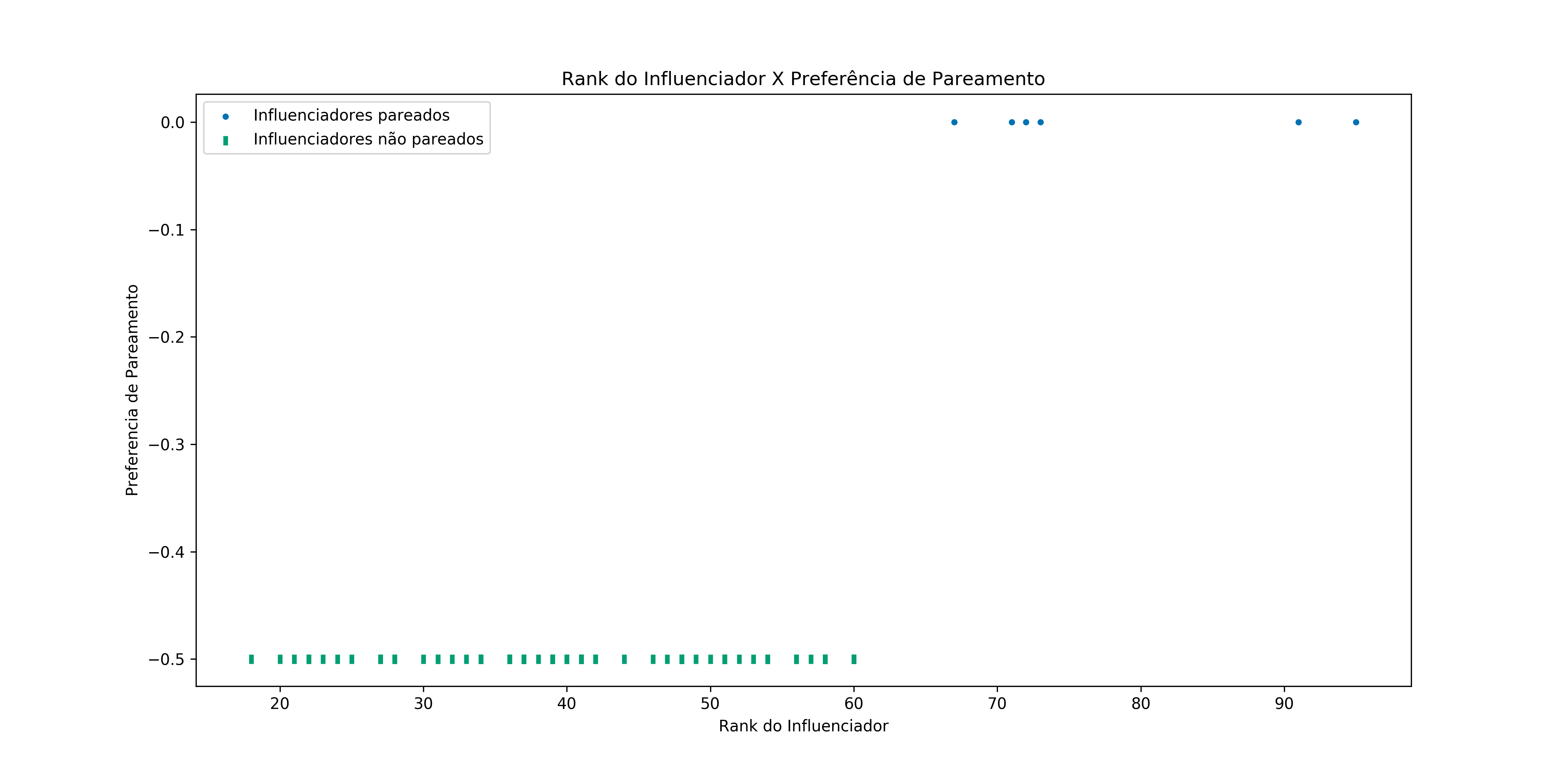}
        \caption*{by \gls{FMC}}
    \end{minipage}
    \caption{Influencer Rank and Preference}\label{fig:img-super-influenciador-preferencia-rank}
\end{figure*}

The consumer classification does not change the fact that it was not possible to pair the consumer with a specific advertisement (see \cref{fig:img-super-influenciador-preferencia-rank}).

\section*{Conclusions}

The ``\textbf{Gale-Shapley}'' algorithm is a mature solution, not only effective in forming optimal pairs but also efficient in its applicability to a real-world problem.

\subsection*{Generic conclusions}


In a general context, this work demonstrates a practical application of the ``\textbf{Gale-Shapley}'' algorithm, and the solution found meets a real need.

\subsection{Generic conclusions}


In particular, this work solves a problem of pairing online media influencers with product and service campaigns from merchants interested in improving their performance in current media.



\printglossary[type=main]
\bigskip
\printglossary[type=acronym]

\nocite{*}
\printbibliography[heading=bibintoc]

\end{document}